\documentclass{article}

\usepackage{arxiv}

\usepackage[utf8]{inputenc} % allow utf-8 input
\usepackage[T1]{fontenc}    % use 8-bit T1 fonts
\usepackage{hyperref}       % hyperlinks
\usepackage{url}            % simple URL typesetting
\usepackage{booktabs}       % professional-quality tables
\usepackage{amsfonts}       % blackboard math symbols
\usepackage{nicefrac}       % compact symbols for 1/2, etc.
\usepackage{microtype}      % microtypography
\usepackage{lipsum}

\usepackage{amsfonts}
\usepackage{amsmath}
\usepackage{subcaption}
\usepackage{textcomp}
\usepackage{graphicx}
\usepackage{amssymb}
\usepackage[inkscapelatex=false]{svg}
\usepackage{mathtools}
\usepackage{newtxtext,newtxmath}
\usepackage{xcolor}
\usepackage{float}
\usepackage{algorithm}
\usepackage{algpseudocode}
\usepackage{stmaryrd}

\newcommand{\R}{{\mathbb{R}}}

\newcommand{\N}{{\mathbb{N}}}

\newcommand{\Wbo}{{\mathcal{W}}}

\newcommand{\X}{{\mathbf{X}}}

\newcommand{\w}{{\mathsf{w}}}
\newcommand{\W}{{\mathbf{W}}}
\newcommand{\U}{{\mathbf{U}}}

\newtheorem{theorem}{Theorem}[section]

\newtheorem{assumption}{Assumption}

\newtheorem{definition}[theorem]{Definition}
\newtheorem{lemma}[theorem]{Lemma}
\newenvironment{proof}{\paragraph{Proof:}}{\hfill$\square$}
\newtheorem{remark}[theorem]{Remark}
\newtheorem{problem}[theorem]{Problem}

\title{Formally Verified Neural Network Controllers for Incremental Input-to-State Stability of Unknown Discrete-Time Systems
\thanks{ This work was supported in part by the ARTPARK.}
}

\author{
 Ahan Basu \\
  Centre for Cyber-Physical Systems\\
  IISc, Bengaluru, India\\
  \texttt{ahanbasu@iisc.ac.in} \\
   \And
 Bhabani Shankar Dey \\
  Centre for Cyber-Physical Systems\\
  IISc, Bengaluru, India\\
  \texttt{bhabanishan1@iisc.ac.in} \\
  \And
 Pushpak Jagtap \\
  Centre for Cyber-Physical Systems\\
  IISc, Bengaluru, India\\
  \texttt{pushpak@iisc.ac.in} \\
}

\begin{document}
\maketitle

\begin{abstract}
    This work aims to synthesize a controller that ensures that an unknown discrete-time system is incrementally input-to-state stable ($\delta$-ISS). In this work, we introduce the notion of $\delta$-ISS control Lyapunov function ($\delta$-ISS-CLF), which, in conjunction with the controller, ensures that the closed-loop system is incrementally ISS. To address the unknown dynamics of the system, we parameterize the controller as well as the $\delta$-ISS-CLF as neural networks and learn them by utilizing the sampled data from the state space of the unknown system. To formally verify the obtained $\delta$-ISS-CLF, we develop a validity condition and incorporate the condition into the training framework to ensure a provable correctness guarantee at the end of the training process. Finally, the usefulness of the proposed approach is proved using multiple case studies.
\end{abstract}

% \begin{IEEEkeywords}
% Incremental Stability, Lyapunov Analysis, Neural-network based control, Scenario Approach.
% \end{IEEEkeywords}

%%%%%%%%%%%%%%%%%%%%%%%%%%%%%%%%%%%%%%%%%%%%%%%%%%%%%%%%%%%%%%%%%%%%%%%%%%%%%%%%
%%--------------------------------NEW SECTION---------------------------------%%
%%%%%%%%%%%%%%%%%%%%%%%%%%%%%%%%%%%%%%%%%%%%%%%%%%%%%%%%%%%%%%%%%%%%%%%%%%%%%%%%

\section{Introduction}\label{sec:intro}
Unlike traditional stability analysis, which emphasizes convergence to an equilibrium or nominal trajectory, incremental stability examines convergence between arbitrary trajectories. Over the years, incremental stability has gained significant attention for its broad applicability, including nonlinear analog circuit modeling \cite{analog}, cyclic feedback system synchronization \cite{cyclic_feedback}, symbolic model development \cite{symbolic1, zamani2017towards, jagtap2020symbolic, jagtap2017quest}, 
% stability of interconnected systems \cite{DD-Stability},
oscillator synchronization \cite{inter_osci}, and complex network analysis \cite{synchComplex}.

In the last few decades, researchers have developed several tools to analyze incremental stability, including contraction analysis \cite{Contraction}, convergent dynamics \cite{Conv_dyn}, and incremental Lyapunov functions \cite{angeli2002lyapunov,DT-ISS_prop, zamani_inc}. These tools have also been extended to analyze the incremental stability of a wide class of systems, such as nonlinear systems \cite{angeli2002lyapunov}, stochastic systems \cite{biemond2018incremental}, hybrid dynamical systems \cite{biemond2018incremental}, time-delayed systems \cite{chaillet2013razumikhin}, and interconnected switched systems \cite{dey2023incremental}. 

In the course of time, numerous efforts have been made to analyze or verify incremental stability. However, designing controllers that enforce incremental stability has been challenging over the years. Most existing approaches for ensuring incremental stability \cite{zamani2011backstepping, zamani2013backstepping, jagtap2017backstepping} rely on the assumption of complete knowledge of the dynamics and structure of the system. {However, in practice, significant model uncertainties can degrade controller performance,  particularly when the assumptions made during design are violated. While classical feedback control offers robustness guarantees to certain perturbations, it may still face limitations under severe uncertainty or high-dimensional nonlinear dynamics. To address such cases, learning-based methods have gained increasing attention.} The authors in \cite{sangeerth2025controller} use the Gaussian process (GP) to collect data from the system and design a backstepping-like controller accordingly under the assumption that the system has a control-affine structure. The newest approach to designing a controller for making a system incrementally ISS \cite{zaker2024controller} also uses a data-driven technique, while the data are collected from the system trajectories. However, this method is only beneficial if the system has a specific polynomial-type structure. {Apart from being limited to specific structures, both these works need the system to be identified first before the control design step, which is challenging in all cases.} 

Neural network-based methods have gained prominence for synthesizing controllers alongside Lyapunov or barrier functions. This approach in some existing literature, such as \cite{formal_nn_Lyapunov, Neural_CBF, tayal2024learning,FV_DD} addresses the issue of safety and stability while eliminating the need for explicit knowledge of the system dynamics. {In \cite{chang2019neural}, learning neural Lyapunov functions for stabilizing controllers is discussed, but verification is
empirical, whereas \cite{yang2024lyapunov} designs neural controllers with Lyapunov stability guarantee for state/output feedback under structural
assumptions.} Leveraging the universal approximation capabilities of neural networks, functions can be directly synthesized. However, a key challenge is providing formal deterministic guarantees, as neural network-based training relies on discrete samples, which cover only a limited part of the continuous state space. {While falsification-based approaches \cite{chang2019neural} have been proposed to mitigate this by adaptively sampling ``difficult” regions, they provide only probabilistic confidence; in contrast, we try to enforce deterministic satisfaction of $\delta$-ISS conditions through constrained optimization. The CEGIS-based approach \cite{abate2020formal} has been a powerful tool as well to verify neural Lyapunov function, but it works in a post-hoc verification scenario, while in this work, we incorporate verification within the training procedure without the need for any other solver.}

\textit{Contributions}: This work proposes a formally verified neural controller synthesis that ensures incremental input-to-state stability for unknown discrete-time systems. Building upon our previous work \cite{basu2025formally}, which verified incremental stability using neural Lyapunov functions, we now focus on controller synthesis that guarantees incremental stability in the closed loop. To this end, we introduce the notion of incremental input-to-state stable control Lyapunov function ($\delta$-ISS-CLF) and prove that its existence under a suitable controller will lead to incremental stability of the closed-loop system. Now, as we use data-driven techniques, we use the compactness assumption, and the controller is trained using the concept of control barrier function, such that the state space becomes forward invariant. First, we formulate the joint synthesis of the controller and Lyapunov function as a robust optimization problem, which is then approximated by a scenario convex program using a finite dataset and considering the controller and Lyapunov function parameterized as neural networks. Following the Lipschitz continuity assumptions, we propose a validity condition that gives a formal guarantee that the trained neural controller can make the system incrementally stable. The proposed method is validated across multiple case studies. 

The key contributions of this paper are: $(i)$ For the first time, we introduce the notion of incremental input-to-state stable control Lyapunov function. $(ii)$ Since the proposed approach relies on data for neural network training, we need to work with compact sets. To address this, we establish conditions for $\delta$-ISS-CLF to ensure incremental input-to-state stability for systems evolving in compact sets. $(iii)$ A novel training framework is proposed to jointly synthesize a controller and a verifiably correct $\delta$-ISS-CLF, both realized as neural networks for unknown discrete-time systems. $(iv)$ Unlike previous works that require fully actuated or control-affine systems \cite{zamani2013backstepping, jagtap2017backstepping}, our approach handles general unknown dynamics.

\section{Preliminaries and Problem Formulation}\label{sec:Prob}

\textit{Notations}: The symbols $\N$, $ \N_0$, $ \R$, $\R^+$, and $\R_0^+$ denote the set of natural, nonnegative integers, real, positive real, and nonnegative real numbers, respectively. 
The vector space of real matrices with $ n $ rows and $ m $ columns is denoted by $\R^{n\times m} $. 
{The set of column vectors with $n$ rows is represented by $ \R^n$.}
The Euclidean norm is represented using $|\cdot |$. 
% The Infinity norm is represented using $\lVert \cdot \rVert$. 
Given a function $\varphi: \N_0 \rightarrow \R^m$, {its sup norm} is given by $\lVert \varphi \rVert = \sup\{|\varphi(k)| : k \in \N_0\}$.
For $a, b \in \N_0$ with $a \leq b$, the closed interval in $\N_0$ is denoted by $[a; b]$.  
A vector $x \in \mathbb{R}^{n}$ with entries $x_1, \ldots, x_n$ is represented as $[x_1, \ldots, x_n]^\top$, where $x_i \in \mathbb{R}$ denotes the $i$-th element of the vector and $i \in [1;n]$.
% A set with elements a,b,c is denoted by \{a,b,c\}. 
{The set of diagonal matrices with non-negative entries} in $\R^{n\times n}$ is denoted by $\mathcal{D}_{\geq 0}^n$.
Given a matrix $M\in\R^{n\times m}$, $M^\top$ represents the transpose of matrix $M$. %Given a matrix $P\in\R^{n\times n}$, $\Tr(P)$ represents trace of matrix $P$. 
A continuous function $\alpha: \R_0^+ \rightarrow \R_0^+$ is said to be class $\mathcal{K}$ if $\alpha(s)>0 $ for all $s>0$, strictly increasing and $\alpha(0)=0$. It is class $\mathcal{K}_\infty$ if it is class $\mathcal{K}$ and $\alpha(s)\rightarrow\infty$ as $s\rightarrow\infty$.
A continuous function $\beta: \R_0^+ \times \R_0^+ \rightarrow \R_0^+$ is said to be a class $\mathcal{KL}$ if $\beta(s,t)$ is a class $\mathcal{K}$ function with respect to $s$ for all $t$ and for fixed $s>0$, $\beta(s,t) \rightarrow 0 $ if $t\rightarrow \infty$. 
{For two functions $\gamma, \kappa \in \mathcal{K}_{\infty}$, we say $\gamma < \kappa$ if $\gamma(s) < \kappa(s)$ for all $s>0$.}
For any compact set $\mathcal{C}, \partial \mathcal{C}$ and $int(\mathcal{C})$ denote the boundary and interior of the set $\mathcal{C}$, respectively.
% The complement of a set $S$ within a set $\mathcal{S}$ is denoted by $\mathcal{S} \backslash S$.

\subsection{Incremental Input-to-State Stability}
Consider the discrete-time control systems as defined next.
\begin{definition}[Discrete-time Control Systems] \label{def:system}
A discrete-time control system (dt-CS) is represented by the tuple $\Xi = (\X, \U, f)$, where $\X \subseteq \R^n$ is the state-space of the system, $\U \subseteq \R^m$ is the input set of the system and $f: \X \times \U \rightarrow \X$ describes the state evolution via the following difference equation:
\begin{equation}\label{eq:act_system}
        \mathsf{x}(k+1) = f(\mathsf{x}(k), \mathsf{u}(k)), \quad \forall k \in \N_0,
\end{equation}
with $\mathsf{x}(k) \in \X$ and $\mathsf{u}(k) \in \U$ are the state and input of the system at $k$-th instance, respectively.
\end{definition}

Now we define the closed-loop system under feedback controller $g$ represented as $\Xi_g = (\X, \W, \U, f, g)$, where $\X \subseteq \R^n$ is the state-space of the system, $\U \subseteq \R^m$ is the internal input set of the system, $\W \subseteq \R^p$ is the external input set of the system, $g:\X \times \W \rightarrow \U$ and $f:\X \times \U \rightarrow \X$  are maps describing the state evolution as:
\begin{equation}\label{eq:system}
        \mathsf{x}(k+1) = f(\mathsf{x}(k), g(\mathsf{x}(k),\mathsf{w}(k))), \quad \forall k \in \N_0,
\end{equation}
where $\mathsf{x}(k) \in \X$ and $\mathsf{w}(k) \in \W$ are the state and external input of the closed-loop system at $k$-th instance, respectively. Note that both $f$ and $g$ are considered to be Lipschitz continuous.
% to guarantee the existence and uniqueness of the solution of \eqref{eq:system}.

Let $\mathsf{x}_{x,\mathsf{w}}(k)$ be the state of the closed-loop system \eqref{eq:system} at time instance $k$ starting from the initial state $x\in\X$ under the sequence of external input $\mathsf{w}$. Next, we define the notion of incremental input-to-state stability ($\delta$-ISS) for the closed-loop system \eqref{eq:system}.

\begin{definition}[$\delta$-ISS \cite{DT-ISS}] \label{def:inc-stable_iss}
    The system $\Xi_g = (\R^n, \R^p, \R^m, f, g)$ in \eqref{eq:system} is said to be incrementally input-to-state stable ($\delta$-ISS) if there exist a class $\mathcal{KL}$ function $\beta$ and a class $\mathcal{K}_\infty$ function $\gamma$, such that for any $k \in \N_0$, for all $x, \hat x \in \X\subseteq \R^n $ and any external input sequence $\mathsf{w},\hat{\mathsf{w}}$ {(where $\mathsf{w}, \hat{\mathsf{w}}: \N \rightarrow \W$)} the following holds:
    \begin{equation}\label{eq:gas-system}
        |\mathsf{x}_{x,\mathsf{w}}(k)-\mathsf{x}_{\hat x,\hat{\mathsf{w}}}(k)| \leq \beta(|x-\hat x|,k) + \gamma(\lVert \mathsf{w} - \hat{\mathsf{w}} \rVert).
    \end{equation}
\end{definition}
If {$\mathsf{w} = \hat{\mathsf{w}} = 0$ and $\X=\R^n$}, one can recover the notion of incremental global asymptotic stability as defined in \cite{DT-ISS_prop}. 

Next, we define the incremental input-to-state stable control Lyapunov function ($\delta$-ISS-CLF) and present the sufficient conditions ensuring incremental stability of the closed-loop system under a $\delta$-ISS-CLF–based controller. 
\begin{definition}\label{def:ISS-Lf_gen}
    The function $V:\R^n \times \R^n \rightarrow {\R_0^+}$ is said to be a $\delta$-ISS control Lyapunov function ($\delta$-ISS-CLF) for closed-loop system $\Xi_g = (\R^n, \R^p, \R^m, f, g)$ in \eqref{eq:system}, if there exist a controller $g: \R^n \times \R^p \rightarrow \R^m$, class $\mathcal{K}_\infty$ functions $\alpha_1, \alpha_2, \alpha_3$, and a class $\mathcal{K}$ function $\sigma$ such that:
    \begin{enumerate}\label{cond:ISS-Lf_gen}
        \item[(i)] for all  $x,\hat{x}\in \R^n$, $\alpha_1(|x-\hat{x}|) \leq V(x,\hat{x}) \leq \alpha_2(|x-\hat{x}|),$
        \item[(ii)] for all $x,\hat{x}\in \R^n$ and for all $w, \hat{w} \in \R^p$,\\ $V(f(x, g(x,w)),f(\hat{x},g(\hat{x},\hat{w}))) - V(x,\hat{x})\leq -\alpha_3(|x - \hat{x}|) + \sigma(|w-\hat{w}|)$.
    \end{enumerate}
\end{definition}  

The following theorem describes $\delta$-ISS in terms of the existence of a $\delta$-ISS control Lyapunov function.

\begin{theorem}\label{th:admit_overall}
The closed-loop discrete-time control system \eqref{eq:system} is said to be incrementally input-to-state stable with respect to input $\w$, if there exists a $\delta$-ISS control Lyapunov function as defined in Definition \ref{def:ISS-Lf_gen}.
\end{theorem}

\begin{proof}
    The proof is similar to that of \cite[Theorem 1]{DT-ISS}.
\end{proof}

In this paper, we aim to present a data-driven approach to tackle the problem of incremental stability. This requires working with compact sets, so we first introduce the notion of $\delta$-ISS-CLF for compact sets. To do this, we begin by revisiting the notion of control forward invariance.
\begin{definition}[Robustly Forward Invariant Set \cite{liu2019compositional}]
    {A set $\X$ is said to be robustly forward invariant with respect to the system \eqref{eq:system} if for every $(x,w) \in \X \times \W$, there exists some control input $u:=g(x,w) \in \U$ such that $f(x, g(x,w)) \in \X$. Now, if there exists a controller $g:\X \times \W \rightarrow \U$ such that the set $\X$ becomes robustly forward invariant, then the controller is said to be a forward invariant controller corresponding to the robustly invariant set $\X$ with respect to external input $\W$}.
\end{definition}

Now we introduce the notion of $\delta$-ISS-CLF for the closed-loop system $\Xi_g$ where the sets $\X \subset \R^n, \W \subset \R^p$ are compact and $\X$ is considered to be a robustly forward invariant under the controller $g$. Then, the definition of $\delta$-ISS-CLF becomes:

\begin{definition}\label{def:ISS-Lf}
    The function ${V:\X \times \X \rightarrow \R_0^+}$ is said to be a $\delta$-ISS control Lyapunov function ($\delta$-ISS-CLF) for closed-loop system $\Xi_g = (\X, \W, \U, f, g)$ in \eqref{eq:system}, where $\X$ and $\W$ are compact sets, if there exist a forward invariant controller $g: \X \times \W \rightarrow \U$, class $\mathcal{K}_\infty$ functions $\alpha_1, \alpha_2, \alpha_3$, and a class $\mathcal{K}$ function $\sigma$ s.t.
    \begin{enumerate}\label{cond:ISS-Lf}
        \item[(i)] for all  $x,\hat{x}\in \X$, $\alpha_1(|x-\hat{x}|) \leq V(x,\hat{x}) \leq \alpha_2(|x-\hat{x}|),$
        \item[(ii)] for all $x,\hat{x}\in \X$ and for all $w, \hat{w} \in \W$,\\ $V(f(x, g(x,w)),f(\hat{x},g(\hat{x},\hat{w}))) - V(x,\hat{x})\leq -\alpha_3(|x - \hat{x}|) + \sigma(|w-\hat{w}|)$.
    \end{enumerate}
\end{definition} 

\begin{theorem}\label{th:admit}
The closed-loop discrete-time control system \eqref{eq:system} is said to be incrementally input-to-state stable within the compact state space $\X$, if there exists a $\delta$-ISS control Lyapunov function under the forward invariant controller $g$ as defined in Definition \ref{def:ISS-Lf}.
\end{theorem}

\begin{proof}
    The proof is presented in Appendix A. 
    % \ref{appendix:admit}.
\end{proof}

% \begin{remark}
%     { Since the neural networks are trained over samples collected from the state space only, it will be difficult to provide a formal guarantee for the unseen points that lie outside the considered domain. Therefore, it is required that the trajectory stays inside the domain, which is ensured by the CBF.}
% \end{remark}

\subsection{Control Barrier Function}
{The notion of CBF is introduced to ensure that the compact set $\X$ becomes robustly forward invariant. Since the $\delta$-ISS-CLF in Definition \ref{def:ISS-Lf} is defined over a compact set, it is essential to ensure that the closed-loop trajectories remain within $\X$ for the validity of the incremental stability guarantees.} We leverage the notion of control barrier function (CBF) introduced next. 

% \begin{definition}[\cite{jagtap2020compositional}]
\begin{definition}[\cite{Neural_CBF}]\label{def:CBC}
    Given a dt-CS $\Xi$ with compact state space $\X$. Let a function $h:\X \rightarrow \R$ is given as 
\begin{subequations}\label{eq:leq_BC}
       \begin{align}
        h(x) &= 0, \hspace{0.3em} \forall x \in \partial \X, \\
        h(x) &< 0, \hspace{0.3em} \forall x \in int(\X).
    \end{align} 
    \end{subequations}
    Then, $h$ is said to be a control barrier function (CBF) for the system $\Xi$ in \eqref{eq:system} if {for every $(x,w) \in \X \times \W$, there exists some control input $u:=g(x,w) \in \U$} such that the following holds:
    \begin{align} \label{eq:diff_BC}
        {h(f(x,g(x,w))) -h(x) \leq 0.}
    \end{align}
    % for some $\kappa \in \mathcal{K}_{\infty}$ with $\kappa \leq \mathcal{I}_d$.
\end{definition}
Now, based on Definition \ref{def:CBC}, {we can design the forward invariant controller $g: \X \times \W \rightarrow \U$ that will make the state-space $\X$ control forward invariant.} The following lemma allows us to synthesize the controller to enforce control invariance.

\begin{lemma}\label{lem:cfi_guarantee}
    Consider a dt-CS $\Xi$ in \eqref{eq:act_system} and let $h:\X \rightarrow \R$ be a control barrier function as defined in Definition \ref{def:CBC}. Then, the controller $g:\X \times \W \rightarrow \U$ satisfying condition \eqref{eq:diff_BC} will make the set $\X$ robustly forward invariant.
    % The set $\X$ will be a robustly forward invariant set for a system $\Xi$ in \eqref{eq:act_system} if there exists a function $h$ that satisfies the conditions of Definition \ref{def:CBC}.
\end{lemma}

\begin{proof}
    Consider a function $h:\X \rightarrow \R$ such that it satisfies condition \eqref{eq:leq_BC}, \textit{i.e.}, $h(x) = 0, \forall x \in \partial \X$ and $h(x) < 0, \forall x \in int(\X)$. Now we assume that there exists a controller $g(x,w)$ such that condition \eqref{eq:diff_BC} is satisfied. Then, one can easily infer $\forall x \in \X, h(f(x,g(x,w))) \leq h(x) \leq 0$ which readily implies $f(x,g(x,w)) \in \X$ which implies that the trajectory of the system will remain within the same set $\X$ under the action of the controller $g$. Hence, the set $\X$ is control forward invariant.
\end{proof}

\subsection{Problem Formulation}
Now, we are ready to discuss the main problem of this paper. This paper considers that discrete-time control systems are unknown; that is, the map $f: \X \times \U \rightarrow \X$ is not known. The main problem of this paper is stated below.
\begin{problem}\label{prob}
    Given a dt-CS $\Xi=(\X,\U,f)$ over compact state-space $\X$, as defined in \eqref{eq:act_system} with unknown dynamics $f$, the main objective is to synthesize a forward invariant feedback controller $g: \X \times \W \rightarrow \U$ that enforces the closed-loop system $\Xi_g=(\X,\W,\U,f,g)$ in \eqref{eq:system} to be $\delta$-ISS with respect to external input $\w$ within state space $\X$.
\end{problem}

Problem \ref{prob} can be reformulated as finding the $\delta$-ISS-CLF function that satisfies the conditions of Definition \ref{def:ISS-Lf} under the existence of some forward invariant controller $g$.

In contrast to previous studies on controller design \cite{zamani2011backstepping}-\cite{zaker2024controller}, which depend on precise knowledge or a specific structure of the system dynamics, our objective is to develop a controller that achieves $\delta$-ISS for the closed-loop system without requiring exact knowledge or a defined structure of the dynamics.

To address the issues in determining the $\delta$-ISS-CLF and the corresponding controller, we present a neural network-based framework that satisfies the conditions of Definition \ref{def:ISS-Lf} and provides a formal guarantee for the obtained neural $\delta$-ISS-CLF and neural controller.
%%%%%%%%%%%%%%%%%%%%%%%%%%%%%%%%%%%%%%%%%%%%%%%%%%%%%%%%%%%%%%%%%%%%%%%%%%%%%%%%
%%--------------------------------NEW SECTION---------------------------------%%
%%%%%%%%%%%%%%%%%%%%%%%%%%%%%%%%%%%%%%%%%%%%%%%%%%%%%%%%%%%%%%%%%%%%%%%%%%%%%%%%
\section{Neural $\delta$-ISS Control Lyapunov Function}\label{sec:Neural Lyapunov}

In this section, we try to find an $\delta$-ISS-CLF, under which the closed-loop system will be $\delta$-ISS according to Theorem \ref{th:admit}. To do so, we first reframe the conditions (i) and (ii) of Definition \ref{def:ISS-Lf} and \eqref{eq:diff_BC} of Definition \ref{def:CBC} as a robust optimization problem (ROP):
\begin{subequations} \label{eq:RCP}
\begin{align}
& \min_{[\eta, d]} \quad \eta  \notag \\
& \textrm{s.t.} \notag \\
& {\forall x,\hat{x} \in \X, x = \hat{x}: V(x,\hat{x}) = 0, }\\
& \forall x,\hat{x} \in \X, x \neq \hat{x}: -V(x,\hat{x}) + \alpha_1(|x-\hat{x}|) \leq \eta, \label{eq:geq} \\
& \forall x,\hat{x} \in \X, x \neq \hat{x}: V(x,\hat{x}) - \alpha_2(|x-\hat{x}|) \leq \eta, \label{eq:leq} \\
& \forall x,\hat{x} \in \X, x \neq \hat{x}, \forall w, \hat{w} \in \W: \notag \\
& \quad V(f(x,g(x, w)),f(\hat{x},g(\hat{x}, \hat{w}))) - V(x,\hat{x}) + \alpha_3(|x-\hat{x}|) - \sigma(|w - \hat{w})|) \leq \eta, \label{eq:diff} \\
& \forall x \in \X,\forall w \in \W:\  {h(f(x,g(x,w))) - h(x)} \leq \eta, \label{eq:diff_BC_ROP} \\
& d = [V, g, \alpha_1, \alpha_2, \alpha_3, \sigma]. \notag
\end{align}
\end{subequations}

Note that $V(\cdot, \cdot) \in \{V|V: \X \times \X \rightarrow\R_0^+\}, g(\cdot, \cdot) \in \{g|g: \X \times \W \rightarrow\U\}, \alpha_1, \alpha_2, \alpha_3 \in \mathcal{K}_{\infty}, \sigma \in \mathcal{K}$. If the optimal solution of ROP $\eta^* \leq 0$, under the constraints defined in \eqref{eq:RCP} leads to the satisfaction of conditions of Definition \ref{def:ISS-Lf} and \ref{def:CBC}, and thus the function $V(x, \hat{x})$ will be a valid $\delta$-ISS-CLF for the unknown system enforcing the system to be $\delta$-ISS under the forward invariant controller $g$.

However, there are several challenges in solving the ROP. They are listed as follows:
\begin{itemize}
    \item[(C1)] The structures of the controller, as well as the $\delta$-ISS-CLF, are unknown. Hence, the solution to the ROP is challenging. 
    \item[(C2)] The function $f$ is unknown, therefore we cannot directly incorporate conditions \eqref{eq:diff} and \eqref{eq:diff_BC_ROP}. So, solving the ROP becomes non-trivial.
    \item[(C3)] The structures of the $\mathcal{K}_{\infty}$ functions $\alpha_1, \alpha_2, \alpha_3$ and class $\mathcal{K}$ function $\sigma$ are unknown.
    \item[(C4)] State space is continuous in nature, so there will be infinitely many equations in the ROP, making the solution of the ROP intractable.
\end{itemize}

To overcome these challenges, the following subsections made some assumptions. To address challenge (C1), we parametrize $\delta$-ISS-CLF and the controller as feed-forward neural networks denoted by $V_{\theta,b}$ and $g_{\Bar{\theta}, \Bar{b}}$, respectively, where $\theta, \Bar{\theta}$ are weight matrices and $b, \Bar{b}$ are bias vectors.
% The detailed structures of these neural networks are discussed in the next subsection.

% \subsection{Neural Network Structure}

\subsection{Construction of SOP and formal verification}\label{sec:formal}

The $\delta$-ISS-CLF neural network consists of an input layer with $2n$ (two times the dimension of the system $n$) and an output layer with one neuron, reflecting the scalar nature of $\delta$-ISS-CLF. The network includes $l_v$ hidden layers with each hidden layer containing $h_v^i, i \in [1;l_v]$ neurons, where both values are arbitrarily chosen.

The activation function of all the layers except the output layer is chosen to be any slope-restricted nonlinear function $\varphi(\cdot)$ (for example, ReLU, Sigmoid, Tanh, etc.). Hence, the resulting neural network function is obtained by recursively applying the activation function as follows:
\[
\begin{cases}
t^0 = [x^\top,\hat{x}^\top]^\top , x,\hat{x} \in \X, \\
t^{i+1} = \phi_i(\theta^it^i + b^i) \hspace{0.2 em} \text{for} \hspace{0.2 em} i \in [0;l_v-1], \\
V_{\theta,b}(x,\hat{x}) = \theta^{l_v}t^{l_v} + b^{l_v}, 
\end{cases}
\]
where $\phi_i:\R^{h_v^i} \rightarrow \R^{h_v^i}$ defined as $ \phi_i(q^i) := [\varphi(q_1^i), \ldots, \varphi(q_{h_v^i}^i)]^\top$. The notation for the controller neural network $g_{\Bar{\theta},\Bar{b}}$ is similar. In this case, the input and output layers have dimensions of $n+p$ and $m$, respectively. The number of hidden layers of the controller neural network is $l_c$ and each layer has $h_c^i, i\in [1;l_c]$ neurons. The activation function for the controller neural network is a similar slope-restricted function. Hence, the resulting controller neural network is given by,

$\begin{cases}
z^0 = [x^\top,w^\top]^\top , x \in \X, w \in \W, \\
z^{i+1} = \phi_i(\bar{\theta}^iz^i + \bar{b}^i) \hspace{0.2 em} \text{for} \hspace{0.2 em} i \in [0;l_c-1], \\
g_{\Bar{\theta},\Bar{b}}(x,w) = \bar{\theta}^{l_c}z^{l_c} + \bar{b}^{l_c}.
% \begin{cases}
%     u_{\min}, \quad \bar{\theta}^{l_c}z^{l_c} + \bar{b}^{l_c} \leq u_{\min}, \\
%     u_{\max}, \quad \bar{\theta}^{l_c}z^{l_c} + \bar{b}^{l_c} \geq u_{\max}, \\
%     \bar{\theta}^{l_c}z^{l_c} + \bar{b}^{l_c}, \quad \text{otherwise}.
% \end{cases}
\end{cases}$

% \subsection{Training Data Generation and Formal Correctness}

Now, to overcome the challenges (C2) and (C3), we raise the following assumptions:
\begin{assumption}\label{assum:black_box}
    We consider having access to the black box or simulator model of the system. Hence, given a state-input pair $(x,u)$, we will be able to generate the next state $f(x,u)$.
\end{assumption}
%To alleviate the challenge (C3), we make another assumption, which is mentioned below.
\begin{assumption}\label{assum:K function}
    We assume that the class $\mathcal{K}_\infty$ functions $\alpha_i, i \in \{1,2,3\}$ are of degree $\gamma_i$ with respect to $|x - \hat{x}|$ and the class $\mathcal{K}$ function $\sigma$ is of degree $\gamma_w$ with respect to $|w - \hat{w}|$, i.e., $\alpha_i(|x - \hat{x}|)=k_i|x - \hat{x}|^{\gamma_i}$, $i\in\{1, 2, 3\}$ and $\sigma(|w - \hat{w}|) = k_w|w - \hat{w}|^{\gamma_w}$, where $k := [k_1, k_2, k_3, k_w]$ and $ \Gamma := [\gamma_1, \gamma_2, \gamma_3, \gamma_w]$ are user-defined parameters. {The choice of $\Gamma$ is made in such a way that these functions become convex.}
\end{assumption}

Now, for training the neural $\delta$-ISS-CLF and the controller, one requires data from the state space and the input space. 
{Also, to overcome the challenge C4, we use a finite number of samples from the compact state space $\X$ and the input space $\W$. To do so, we collect $N$ samples $x_s$ from the state space $\X$ where $s \in [1;N]$. We consider a ball $B_{\varepsilon_x}(x_s)$ around each sample $x_s$ with radius $\varepsilon_x$ such that for all $x \in \X$, there exists an $x_s$ such that $|x - x_s|\leq \varepsilon_x$. This ensures that,  $\bigcup_{s=1}^{N} B_{\varepsilon_x}(x_s) \supset \X$. A similar strategy has been applied to collect $M$ samples $w_p$ from the input space $\W$, where the radius of the ball is $\varepsilon_u$.}

% We collect samples $x_s$ and $w_p$ from $\X$ and $\W$, where $s = [1;N], p=[1;M]$. Consider ball $B_{\varepsilon_x}(x_s)$ and $B_{\varepsilon_u}(w_p)$ around each sample $x_s$ and $w_p$, with radius $\varepsilon_x$ and $\varepsilon_u$, such that, $\X \subseteq \bigcup_{s=1}^{N} B_{\varepsilon_x}(x_s)$ and $\W \subseteq \bigcup_{p=1}^{M} B_{\varepsilon_u}(w_p)$ with :
% \begin{subequations}\label{eq:ball}
%     \begin{align}
%         |x - x_s| &\leq \varepsilon_x , \forall x \in B_{\varepsilon_x}(x_s), \\
%         |w - w_p| &\leq \varepsilon_u , \forall w \in B_{\varepsilon_u}(w_p).
%     \end{align}
% \end{subequations}

We consider $\varepsilon = \max(\varepsilon_x, \varepsilon_u)$.  Collecting the sampled points $x_s$ and $w_p$, we form the data sets denoted by:
\begin{align}\label{set:SCP}
\mathcal{X} \!=\! \{x_s | \bigcup_{s=1}^{N} B_{\varepsilon_x}(x_s) \!\supset\! \X \},\ 
\Wbo \!= \!\{w_p | \bigcup_{p=1}^{M} B_{\varepsilon_u}(w_p) \!\supset\! \W \}.
\end{align}

Next, we construct a scenario convex optimization problem (SCP) to address the challenge (C4) related to the ROP, as defined in \eqref{eq:RCP}. This problem is constructed using the data sets and the assumptions in \ref{assum:black_box} and \ref{assum:K function}, with finite constraints.
% \cite{Neural_CBF, dd-stt}:
\begin{subequations} \label{eq:SCP}
\begin{align}
& \min_{\eta} \quad \eta  \notag \\
& \textrm{s.t.} \notag \\
& {\forall x_q,x_r \in \mathcal{X}, x_q = x_r: V_{\theta,b}(x_q,x_r) = 0, }\label{eq:eq_SOP}\\
& \forall x_q,x_r \in \mathcal{X}, x_q \neq x_r: -V_{\theta,b}(x_q,x_r)\! + \!k_1|x_q-x_r|^{\gamma_1} \!\leq \!\eta, \label{eq:geq_SOP} \\
& \forall x_q,x_r \in \mathcal{X}, x_q \neq x_r: V_{\theta,b}(x_q,x_r) - k_2|x_q-x_r|^{\gamma_2} \leq \eta, \label{eq:leq_SOP} \\
& \forall x_q,x_r \in \mathcal{X}, x_q \neq x_r,  \forall w_q, w_r \in \Wbo: \notag \\
& \quad V_{\theta,b}(f(x_q,g_{\Bar{\theta},\Bar{b}}(x_q, w_q)),f(x_r,g_{\Bar{\theta},\Bar{b}}(x_r, w_r))) -  V_{\theta,b}(x_q,x_r) + k_3 |x_q - x_r|^{\gamma_3} - k_w|w_q - w_r|^{\gamma_w}\leq \eta,\label{eq:diff_SOP}  \\
& \forall x_q \!\in\! \mathcal{X}, w_q \in \Wbo: {h(f(x_q,g_{\Bar{\theta},\Bar{b}}(x_q,w_q)))\! -\! h(x_q)} \!\leq\! \eta. \label{eq:diff_BC_SOP}
\end{align}
\end{subequations}

{Note that the optimization program becomes convex as all the constraints, as well as the objective function, are convex with respect to the decision variable. Now,} given the finite number of data samples, the SCP involves a finite set of constraints, making its solution computationally tractable. Let $\eta_S^*$ denote the optimal solution of the SCP. To demonstrate that the solution of the SCP is also a feasible solution for the proposed ROP, we impose the following assumptions regarding Lipschitz continuity:

\begin{assumption}\label{assum:Lipschitz_fun}
    The function $f$ in \eqref{eq:act_system} is Lipschitz continuous with respect to $x$ and $u$ over the state space $\X$ and input space $\U$ with the Lipschitz constants $\mathcal{L}_x$ and $\mathcal{L}_u$. 
\end{assumption}
{We assume the Lipschitz constants of the dynamics are known, even if the system is unknown.} Additionally, these constants can also be estimated following a similar procedure in \cite[Algorithm 2]{FV_DD}. 
% {It is worth mentioning that since the Lipschitz constants are estimated using data samples collected from the state-space, it might not be accurate \cite[Remark 8.2]{FV_DD}, but picking $\alpha$ very small and $\bar{N},M$ very big, we have a good approximation for the Lipschitz constants of the system.}

\begin{assumption}\label{assum:Lipschitz_net}
    We assume that the candidate neural $\delta$-ISS-CLF is Lipschitz continuous with Lipschitz bound $\mathcal{L}_L$ with respect to $(x,\hat{x})$ over the set $\X$. Similarly, the controller neural network has a Lipschitz bound $\mathcal{L}_C$.
\end{assumption}
In the next subsection, we explain how $\mathcal{L}_L$ and $\mathcal{L}_C$ are enforced during the training procedure.

\begin{remark}
    {The class $\mathcal{K}_\infty$ functions and the class $\mathcal{K}$ function of Definition \ref{def:ISS-Lf} are Lipschitz continuous with Lipschitz constants $\mathcal{L}_1, \mathcal{L}_2, \mathcal{L}_3$, and $\mathcal{L}_w$, respectively by virtue of Assumption \ref{assum:K function}, as it is evident that convex continuous functions are Lipschitz continuous inside a compact set. One can estimate the values using the values of $k$ and $\Gamma$.} In addition, the Lipschitz constant $\mathcal{L}_h$ of the function $h$ is already predefined due to the known structure of $h$.
\end{remark}

Under Assumptions \ref{assum:Lipschitz_fun} and \ref{assum:Lipschitz_net}, the following theorem outlines the connection of the solution of SCP \eqref{eq:SCP} to that of ROP \eqref{eq:RCP}, providing a formal {deterministic} guarantee to the obtained $\delta$-ISS-CLF satisfying the incremental stability conditions under the controller $g_{\Bar{\theta},\Bar{b}}$.

\begin{theorem}\label{th:constr}
    Consider a dt-CS $\Xi$ given in \eqref{eq:act_system}. Let $V_{\theta,b}$ be the neural $\delta$ISS control Lyapunov function and let $g_{\Bar{\theta},\Bar{b}}$ be the corresponding controller. Let the optimal value of SCP \eqref{eq:SCP}, $\eta_S^*$, be obtained using data samples collected from the state-space and input space. Then under Assumptions \ref{assum:Lipschitz_fun} and \ref{assum:Lipschitz_net}, if
    \begin{align}\label{eq:cond}
        \eta_S^* + \mathcal{L}\epsilon \leq 0, 
    \end{align}
    where $\mathcal{L} = \max\{\sqrt{2}\mathcal{L}_L + 2\mathcal{L}_1, \sqrt{2}\mathcal{L}_L + 2\mathcal{L}_2, \sqrt{2}\mathcal{L}_L (\mathcal{L}_x + \sqrt{2}\mathcal{L}_u\mathcal{L}_C + 1)+ 2(\mathcal{L}_3+\mathcal{L}_w), \mathcal{L}_h (\mathcal{L}_x + \sqrt{2}\mathcal{L}_u \mathcal{L}_C + 1)\}$, then the obtained $\delta$-ISS control Lyapunov function upon solving the SCP ensures that the closed-loop system is incrementally input-to-state stable under the action of the controller $g_{\Bar{\theta},\Bar{b}}$ {with a deterministic correctness guarantee}.
\end{theorem}

\begin{proof}
The proof is presented in Appendix B.
% of \cite{basu2025controller}. 
\end{proof}

\begin{remark}
{In this work, although we use a scenario-based sampling approach to approximate the ROP, our samples are generated deterministically, unlike the random sampling in \cite{esfahani2014performance} that yields probabilistic guarantees. This deterministic sampling, together with the Lipschitz-based validity condition, ensures the deterministic correctness of the synthesized controller and $\delta$-ISS-CLF.}
\end{remark}

\subsection{Formulation of Loss functions and Training procedure}
Here, we utilize the derived condition \eqref{eq:cond} from Section \ref{sec:formal} and propose a training framework to synthesize provably correct $\delta$-ISS-CLF and the controller parametrized as neural networks. In particular, we train $\delta$-ISS-CLF and controller simultaneously to achieve formal guarantees on their validity by constructing suitable loss functions that incorporate the satisfaction of conditions \eqref{eq:geq}-\eqref{eq:diff_BC_ROP} and \eqref{eq:cond} over the state-space and input-space. 

We consider \eqref{eq:eq_SOP}-\eqref{eq:diff_BC_SOP} as sub-loss functions to construct the actual loss function. The sub-loss functions are:
\begin{subequations}\label{eq:loss_LR}
    \begin{align}
        L_0(\psi,\eta) &= \sum_{x = \hat{x}} \max\big(0,V_{\theta,b}(x,\hat{x})\big ), \\
        L_1(\psi,\eta) &= \sum_{x \neq \hat{x}}\max\big(0,(-V_{\theta,b}(x,\hat{x}) + k_1|x- \hat{x}|^{\gamma_1} - \eta)\big), \\
        L_2(\psi,\eta) &= \sum_{x \neq \hat{x}}\max\big(0,(V_{\theta,b}(x,\hat{x}) - k_2|x- \hat{x}|^{\gamma_2} - \eta)\big), \\
        L_3(\psi,\eta) &= \sum_{x \neq \hat{x}}\max\big(0,(V_{\theta,b}(f(x, g_{\bar{\theta}, \bar{b}}(x,w)), f(\hat{x},g_{\bar{\theta}, \bar{b}}(\hat{x},\hat{w}))) - V_{\theta,b}(x,\hat{x}) + k_3|x - \hat{x}|^{\gamma_3} - k_w|w - \hat{w}|^{\gamma_w} - \eta)\big), \\
        L_4(\psi,\eta) &= \sum \max \big(0, h(f(x,g_{\Bar{\theta},\Bar{b}}(x,w))) - k_hh(x) - \eta\big),
    \end{align}
\end{subequations}
where, $x, \hat{x} \in \mathcal{X}, w , \hat{w} \in \mathcal{W}$, with $\psi = [\theta,b,\Bar{\theta},\Bar{b}]$ and $\eta$ being the trainable parameters. As mentioned, the actual loss function is a weighted sum of the sub-loss functions and is denoted by
\begin{align}
    L(\psi, \eta) &= c_0L_0(\psi, \eta) + c_1L_1(\psi, \eta) + c_2L_2(\psi, \eta) + c_3L_3(\psi, \eta) + c_4L_4(\psi, \eta),
\end{align}
where $c_0,c_1,c_2,c_3,c_4 \in \R^+$ are the weights of the sub-loss functions $L_0(\psi, \eta),L_1(\psi, \eta), L_2(\psi, \eta), L_3(\psi, \eta)$ and $L_4(\psi, \eta)$ respectively. 

Next, the Lipschitz continuity of the candidate neural networks following Assumption \ref{assum:Lipschitz_net} needs to be satisfied. Since neural network candidates consist of {Lipschitz} activation layers, the assumption is already satisfied. 
% Note that, $V_{\theta,b}$ is Lipschitz continuous with a bound $\mathcal{L}_L$, so:
% \begin{align}
%     &\forall a,b,c,d \in \R^n: \notag \\
%     &|V_{\theta,b}(a,b) - V_{\theta,b}(c,d)| \leq \mathcal{L}_L|(a,b) - (c,d)|.
% \end{align}
Now, the following lemma addresses the guarantee of the Lipschitz bound over the training procedure.

\begin{lemma}[\cite{pauli2022b}]
    Suppose $f_{\theta}$ is a $p$-layered feed-forward neural network with $\theta$ as the trainable parameter. Then the neural network is said to be Lipschitz bounded with the Lipschitz constant $\mathcal{L}_N$ if the following semi-definite constraint $\mathcal{M}(\theta,\Lambda)$:=
    \begin{align}\label{eq:mat_ineq}
        \begin{bmatrix}
            A \\ B
        \end{bmatrix}^\top\!
        \begin{bmatrix}
            2 \alpha \beta \Lambda &\!\!\! -(\alpha\!+\!\beta)\Lambda \\ -(\alpha\!+\!\beta)\Lambda & 2\Lambda
        \end{bmatrix}\!
        \begin{bmatrix}
            A \\ B
        \end{bmatrix} 
        \!+\! \begin{bmatrix}
            \mathcal{L}_N^2\textbf{I} & 0 & 0 & 0 \\ 0 & 0 & 0 & 0 \\ 0 & 0 & 0 & -\theta_p^\top \\ 0 & 0 & -\theta_p & \textbf{I} 
        \end{bmatrix}\! \geq \!0,
    \end{align}
    holds, where 
    $A = \begin{bmatrix}
        \theta_0 & \ldots & 0 & 0 \\ \vdots & \ddots & \vdots & \vdots \\0 & \ldots & \theta_{p-1} & 0  
    \end{bmatrix}, B = \begin{bmatrix}
        0 & \textbf{I}
    \end{bmatrix}$, $\theta_0, \ldots, \theta_p$ are the weights of the neural network, $\Lambda \in \mathcal{D}_{\geq 0}^{n_i}, i \in \{1, \ldots, p\}$, where $n_i$ denotes number of neurons in $i$-th layer, and $\alpha,$ and $ \beta$ are the minimum and maximum slope of the activation functions, respectively.
\end{lemma} 

We denote the matrix inequalities satisfying the Lipschitz conditions for the $\delta$-ISS-CLF and the controller neural network as $\mathcal{M}(\theta,\Lambda)$ and $\mathcal{M}(\Bar{\theta},\Bar{\Lambda})$, respectively. So, for the satisfaction of conditions \eqref{eq:cond} and \eqref{eq:mat_ineq}, we consider two more loss functions denoted by
\begin{align}
    L_{\mathcal{M}}(\psi,\Lambda,\Bar{\Lambda}) &\!=\! -c_{l_1}\log\det(\mathcal{M}(\theta,\Lambda))\! -\! c_{l_2}\log\det(\mathcal{M}(\Bar{\theta},\Bar{\Lambda})), \label{eq:loss_ineq}\\
    L_v(\eta) &= \max\big(0,(\mathcal{L}\epsilon + \eta)\big), \label{eq:loss_th}
\end{align}
where $\psi,\Lambda,\Bar{\Lambda}$ are trainable parameters.  $\mathcal{L}_L$ and $\mathcal{L}_C$ that appear in $\mathcal{M}(\theta,\Lambda)$ and $\mathcal{M}(\Bar{\theta},\Bar{\Lambda})$ are used to compute $\mathcal{L}$ as described in Theorem \ref{th:constr}. The Lipschitz bounds and the $\varepsilon$ are hyper-parameters chosen a priori. Additionally, $c_{l_1},c_{l_2} \in \R^+$. 

Now, under the trained neural networks corresponding to $\delta$-ISS-CLF and the controller, the following theorem provides a formal guarantee for the closed-loop system to be incrementally input-to-state stable under the action of the controller $g_{\bar \theta,\bar b}$.

\begin{theorem}\label{th:guarantee}
    {Consider a dt-CS $\Xi$ with compact state-space $\X$ and unknown dynamics $f$. Given a compact input space $\W$, we train two neural networks $V_{\theta,b}$ and $g_{\Bar{\theta},\Bar{b}}$ representing the $\delta$-ISS CLF and the controller for the closed-loop system $\Xi_g$,} such that the loss functions converge as $L(\psi, \eta) = 0, L_v(\eta) = 0$ and $L_{\mathcal{M}}(\psi,\Lambda, \bar{\Lambda}) \leq 0$ over the training data sets $\mathcal{X}$ and $\Wbo$. Then, the closed-loop system under the influence of the controller $g_{\Bar{\theta},\Bar{b}}$ is guaranteed to be $\delta$-ISS as defined in Definition \ref{def:inc-stable_iss}.
\end{theorem}

\begin{proof}
{The first loss $L(\psi, \eta) = 0$ implies that the SCP has been solved with optimal $\eta_S^*$, ensuring the closed-loop system to be $\delta$-ISS with $x \in \mathcal{X}$. The second loss $L_v(\eta) = 0$ implies the satisfaction of Theorem \ref{th:constr}, ensuring $\delta$ -ISS for any initial state in $\X$}. The third loss $L_{\mathcal{M}}(\psi,\Lambda, \bar{\Lambda}) \leq 0$ implies the neural networks adhere to the Lipschitz bounds, satisfying Assumption \ref{assum:Lipschitz_net}.
% The proof can be staged over satisfying conditions for three loss functions. 
% \begin{itemize}
%     \item The first loss $L(\psi, \eta) = 0$ implies that the SCP has been solved with the minimum value being $\eta_S^*$. Hence, the controller is trained to ensure that the closed-loop system is $\delta$-ISS with initial conditions belong to $\mathcal{X}$ (that is, finite over data).
%     \item The second loss $L_v(\eta) = 0$ implies the satisfaction of Theorem \ref{th:constr}, which means that the solution of SCP is also valid for ROP. Hence, the closed-loop system is $\delta$ -ISS for any initial state in $\X$ and any external input sequence in $\w$.
%     \item The third loss $L_{\mathcal{M}}(\psi,\Lambda, \bar{\Lambda}) \leq 0$ implies that the matrices $\mathcal{M}(\theta,\Lambda)$ and $\mathcal{M}(\Bar{\theta},\Bar{\Lambda})$ are positive definite, satisfying the Lipschitz continuity assumption of the neural networks with the assumed Lipschitz bounds.
% \end{itemize}
Hence, the satisfaction of the above theorem leads to ensuring that the closed-loop system is $\delta$-ISS under the action of the controller. This completes the proof.
\end{proof}

The training process of the neural $\delta$-ISS-CLF and the corresponding controller is described {in Algorithm \ref{algo:NN_training}}.

\begin{algorithm}
\caption{Training of the Neural Networks}
\label{algo:NN_training}
{\begin{algorithmic}[1]
    \Require Black box model of the system $f(x,u)$, Data sets: $\mathcal{X}, \mathcal{W}$ 
    \Ensure $V_{\theta,b}, g_{\bar{\theta}, \bar{b}}, \eta $
    \State Select the hyperparameters $\epsilon$, $c = [c_0, c_1, c_2, c_3, c_{l1}, c_{l2}], k = [k_1, k_2, k_3, k_w], \Gamma = [\gamma_1, \gamma_2, \gamma_3, \gamma_w], \mathcal{L}_L, \mathcal{L}_C, \mathcal{L}_h$ and number of epochs.
    \State Computation of Lipschitz constants $\mathcal{L}_1, \mathcal{L}_2, \mathcal{L}_3, \mathcal{L}_w$.
    \State Estimation of Lipschitz constants $\mathcal{L}_x$ and $\mathcal{L}_u$ \cite{Lipschitz}. Compute $\mathcal{L}$ using Theorem \ref{th:constr}.
    \State Initialize Neural networks and trainable parameters $\theta,b,\bar{\theta}, \bar{b},\Lambda,\bar{\Lambda},\eta$.
    \For{$i\leq Epochs$ (Training starts here)}
        \State Create batches of training data from $\mathcal{X}, \mathcal{W}$
        \State Find batch losses using \eqref{eq:loss_LR}, \eqref{eq:loss_ineq} and \eqref{eq:loss_th}.
        \State Use ADAM or SGD optimizer with specified learning rate \cite{ruder2016} to reduce loss and update the trainable parameters.
        \If{Theorem \ref{th:guarantee} is satisfied}
            \State \textbf{break}
        \EndIf
    \EndFor
    \State \textbf{return} $V_{\theta,b}, g_{\bar{\theta}, \bar{b}}, \eta$
\end{algorithmic}}
\end{algorithm}

\begin{remark}
    {Note that if the algorithm does not converge successfully, one cannot judge the $\delta$-ISS of the closed-loop system with the specified hyperparameters $c,k,\Gamma, \mathcal{L}_L,\mathcal{L}_C$. To improve convergence, one can incorporate several strategies like reducing the discretization parameter $\epsilon$ \cite{zhao2021learning} or adjusting neural network hyperparameters (architecture, learning rate) \cite{nn_lr} or reconsidering the initial choice of hyperparameters of the algorithm as well.
    % and reconsider the choice of initial parameters. In addition, the number of samples also plays a crucial role in the convergence algorithm. A coarsely sampled state-input space may lead to inadequate samples giving low $\eta$ values and the algorithm might not converge. Therefore, it may be necessary to perform denser sampling to ensure the convergence of the algorithm to enforce the system to be $\delta$-ISS.
    Also, the theoretically ideal case of the convergence of the algorithm is that the loss $L(\psi, \eta)$ should be exactly zero; however, in practice, this is generally infeasible. One can consider that the algorithm has converged based on some very small residual margin in the order of $10^{-6}$ to $10^{-4}$ during implementation.} 
\end{remark}

\begin{remark}
    In addition, the initial feasibility of condition \eqref{eq:mat_ineq} is required to satisfy the criterion of loss {$L_\mathcal{M}$} in \eqref{eq:loss_ineq} according to Theorem \ref{th:guarantee}. Choosing small initial weights and biases for neurons can ensure this condition. {The choice of Lipschitz bounds of the networks is generally kept small, so that the networks are robust enough corresponding to the inputs.}
\end{remark}

% \begin{remark}[Computation of functions $\beta$ and $\gamma$ in \eqref{eq:gas-system}]
%     The class $\mathcal{KL}$ function $\beta$ and the class $\mathcal{K}_\infty$ function $\gamma$ of Definition \ref{def:inc-stable_iss} can be computed using a similar procedure once the trained $\delta$-ISS-CLF is obtained as referred to in \cite{DT-ISS}. The methodology of obtaining the functions will follow the proof mentioned in Appendix \ref{appendix:admit}.
% \end{remark}

\begin{remark}[Dealing with input constraints]
    To keep the output of the controller neural network bounded within the input constraints, one can consider the HardTanh function as the activation function of the last layer of the controller neural network. In that case, $\U$ is assumed to be a polytopic set with bounds given as $u_{\min} \leq u \leq u_{\max}, u \in \U$. Then, the resulting controller will be:
    $\begin{cases}
    z^0 = [x^\top,w^\top]^\top , x \in \X, w \in \W, \\
    z^{i+1} = \phi_i(\bar{\theta}^iz^i + \bar{b}^i) \hspace{0.2 em} \text{for} \hspace{0.2 em} i \in [0;l_c-1], \\
    g_{\Bar{\theta},\Bar{b}}(x,w) =
    \begin{cases}
        u_{\min}, \quad \bar{\theta}^{l_c}z^{l_c} + \bar{b}^{l_c} \leq u_{\min}, \\
        u_{\max}, \quad \bar{\theta}^{l_c}z^{l_c} + \bar{b}^{l_c} \geq u_{\max}, \\
        \bar{\theta}^{l_c}z^{l_c} + \bar{b}^{l_c}, \quad \text{otherwise}.
    \end{cases}
    \end{cases}$
\end{remark}

%%%%%%%%%%%%%%%%%%%%%%%%%%%%%%%%%%%%%%%%%%%%%%%%%%%%%%%%%%%%%%%%%%%%%%%%%%%%%%%%
%%--------------------------------NEW SECTION---------------------------------%%
%%%%%%%%%%%%%%%%%%%%%%%%%%%%%%%%%%%%%%%%%%%%%%%%%%%%%%%%%%%%%%%%%%%%%%%%%%%%%%%%
\section{Case Studies}\label{sec:cases}
The proposed procedure to make a system incremental input-to-state stable using a neural network-based controller is shown through the following case studies of a general nonlinear system and the motion of a rotating rigid spacecraft model. All the case studies were performed using PyTorch in Python 3.10 on a machine with a Windows operating system with Intel Core i7-14700 CPU, 32 GB RAM and NVIDIA GeForce RTX 3080 Ti GPU.

\begin{figure*}[t]
    \centering
    \begin{subfigure}{0.24\textwidth}
        \centering
        \includegraphics[width=0.9\textwidth]{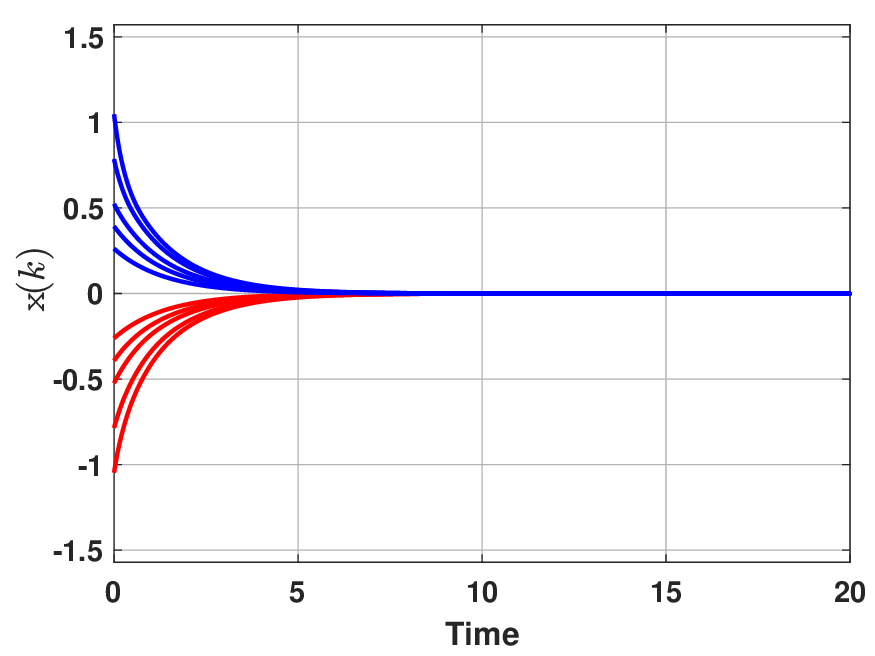}
        \caption{}
    \end{subfigure}
    \hfill
    \begin{subfigure}
    {0.24\textwidth}
        \centering
        \includegraphics[width=0.9\textwidth]{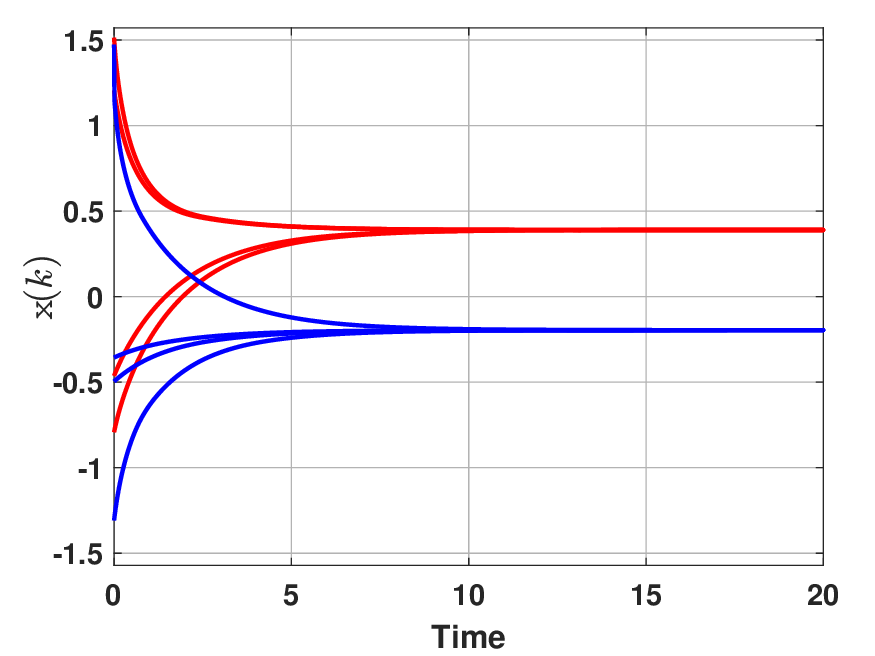}
        \caption{}
    \end{subfigure}
    \hfill 
    \begin{subfigure}{0.24\textwidth}
        \centering
        \includegraphics[width=0.9\textwidth]{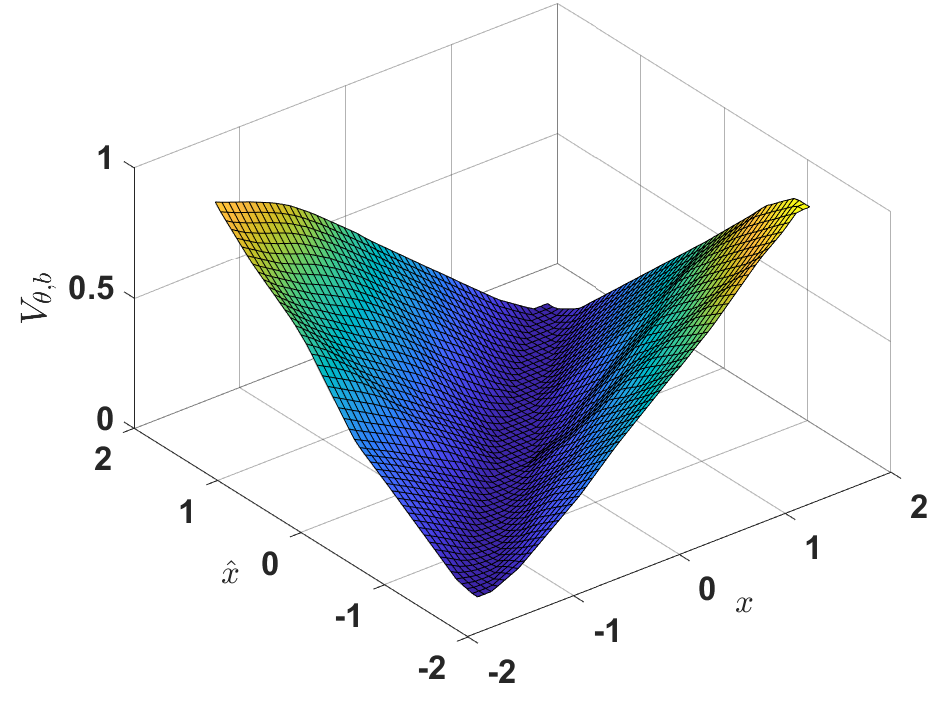}
        \caption{}
    \end{subfigure}
    \hfill
    \begin{subfigure}{0.24\textwidth}
        \centering
        \includegraphics[width=0.9\textwidth]{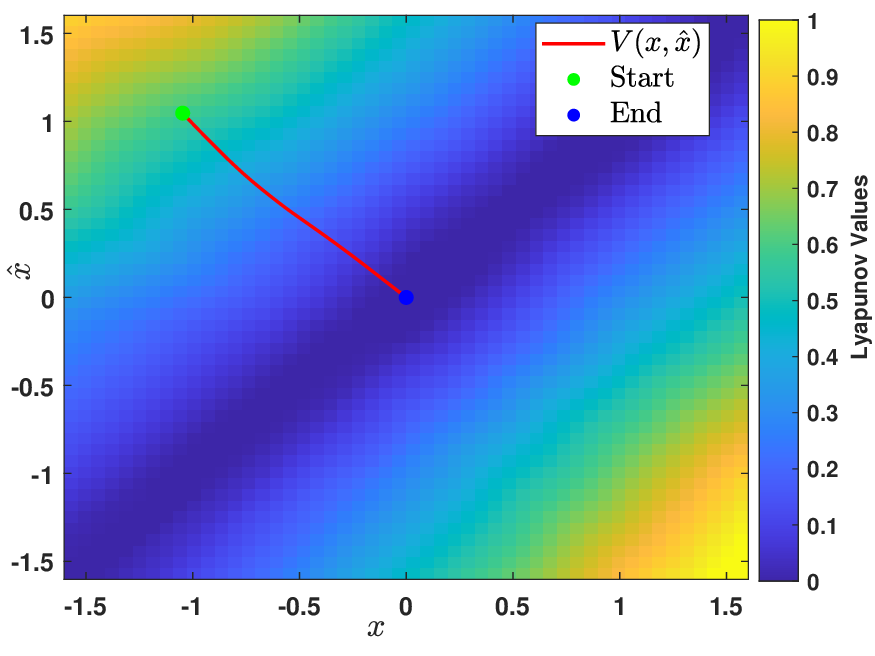}
        \caption{}
    \end{subfigure}
    \vspace{-0.1cm}
    \caption{(a){Trajectories starting from different initial conditions under same input signals, (b) Trajectories starting from different initial conditions under different input signals ($\w(k) = 0.2$ for red curves, $\w(k) = -0.1$ for blue curves) (c) The $\delta$-ISS-CLF plot is greater than zero for all $(x, \hat{x}) \in \X \times \X, x \neq \hat{x}$ and equal to zero while $x = \hat{x}$, (d) The Lyapunov value over the trajectories is decaying with time with the heatmap showing Lyapunov level sets.}}
    \label{fig:sim1}
\end{figure*}

\subsection{Simple Nonlinear System}
We consider a simple nonlinear scalar system, whose discrete-time dynamics is given by:
\begin{align}
    \mathsf{x}(k+1) = \mathsf{x}(k) + \tau(a \sin{(\mathsf{x}(k))} + \tan(\mathsf{u}(k))),
\end{align}
where $x(k)$ denotes the state of the system at $k$-th instant. The constant $a = 0.1$ represents the rate constant of the system. $\tau = 0.01$ is the sampling time. We consider the state space of the system to be $\X = [-\frac{\pi}{2}, \frac{\pi}{2}]$. Moreover, the input set is given as $\W = [-1,1]$. We consider the model to be unknown. { However, the Lipschitz constants $\mathcal{L}_x = 1, \mathcal{L}_u = 0.01$ are known for the system.}
% we estimate the Lipschitz constants $\mathcal{L}_x = 1, \mathcal{L}_u = 0.01$ by picking the parameters $\alpha=0.0005, \bar{N}=1000,M=100$ of \cite[Algorithm 2]{FV_DD}}. 

The goal is to synthesize a controller to enforce the system to be $\delta$-ISS. So, we are to synthesize a valid $\delta$-ISS-CLF $V_{\theta,b}$ under the action of the controller $g_{\Bar{\theta},\Bar{b}}$. To do this, we first fix the training hyper-parameters as $\epsilon = 0.00039, \mathcal{L}_L = 1, \mathcal{L}_C = 20, {k = [0.00001, 0.5, 0.0001, 0.01], \Gamma = [2,2,2,2]}$. So, the Lipschitz constant, $\mathcal{L}$, according to Theorem \ref{th:constr} is $3.25$. We fix the structure of $V_{\theta,b}$ as {$l_v = 1, h_v^1 = 40$} and $g_{\Bar{\theta},\Bar{b}}$ as $l_c = 1, h_c^1 = 15$. {The activation function for both $\delta$-ISS-CLF and the controller is considered to be `ReLU'.}

Now we consider the training data obtained from \eqref{set:SCP} and perform training to minimize the loss functions $L, L_{\mathcal{M}} $ and $ L_v$. The training algorithm converges to obtain $\delta$-ISS-CLF $V_{\theta,b}$ along with $\eta = -0.0015$. Hence, $\eta+\mathcal{L}\epsilon = -0.0015 + 3.25\times0.00039 = -0.00023$, thus by utilizing Theorem \ref{th:guarantee}, we can guarantee that the obtained Lyapunov function $V_{\theta,b}$ is valid and the closed-loop system is guaranteed to be incrementally input-to-state stable under the influence of the controller $g_{\Bar{\theta},\Bar{b}}$.

The successful runs of the algorithm {including the dataset generation time} have an average convergence time of 45 minutes. 

{As can be seen in Figure \ref{fig:sim1}(a), under the same input conditions, the trajectories starting from different initial conditions merge towards each other asymptotically under the influence of the controller.} The $\delta$-ISS-CLF plot for this case is shown in Figure \ref{fig:sim1}(b). Note that, in Figure \ref{fig:sim1}(c), {The Lyapunov level sets are shown using a heatmap plot, while in the same figure, it is shown that the Lyapunov trajectory is decaying over time.}

\subsection{One-Link Manipulator}
We consider a single link manipulator dynamics \cite{lewis2020neural, murray2017mathematical}, whose discrete-time dynamics is given as:
\begin{align}
    \mathsf{x}_1(k+1) &= \mathsf{x}_1(k) + \tau (\mathsf{x}_2(k)),\notag \\
    \mathsf{x}_2(k+1) &= \mathsf{x}_2(k) + \tau \Big(\frac{1}{M}(\mathsf{u}(k) - b\mathsf{x}_2(k))\Big),
\end{align}
where $\mathsf{x}_1,\mathsf{x}_2$ denotes angular position and velocity respectively. The constants $M = 1, b = 0.1$ stand for the mass and damping coefficient of the system, respectively, while $\tau = 0.01$ is the sampling time. We consider the state space of the system to be $\X = [-\frac{\pi}{4}, \frac{\pi}{4}]\times[-\frac{\pi}{4}, \frac{\pi}{4}]$. Moreover, we consider the input set to be bounded within $\W = [-0.5,0.5]$. Also, we consider the model to be unknown. However, the Lipschitz constants $\mathcal{L}_x = 1.01, \mathcal{L}_u = 0.01$ are known for the system. 

The goal is to synthesize a controller to enforce the system to be $\delta$-ISS. To do this, we first fix the training hyper-parameters as $\epsilon = 0.0157, \mathcal{L}_L = 1, \mathcal{L}_C = 40, k_1 = 0.00001, k_2 = 0.5, k_3 = 0.0001, k_w = 0.01$. So, the Lipschitz constant according to Theorem \ref{th:constr} is $3.663$. We fix the structure of $V_{\theta,b}$ as {$l_v = 1, h_v^1 = 40$} and $g_{\Bar{\theta},\Bar{b}}$ as $l_c = 1, h_c^1 = 15$. The training algorithm converges to obtain $\delta$-ISS-CLF $V_{\theta,b}$ along with $\eta = -0.0579$. Hence, $\eta+\mathcal{L}\epsilon = -0.0579 + 3.663\times0.0157 = -0.00039$, thus by utilizing Theorem \ref{th:guarantee}, we can guarantee that the obtained $\delta$-ISS-CLF $V_{\theta,b}$ is valid and that the closed-loop system is assured to be incrementally input-to-state stable under the influence of the controller $g_{\Bar{\theta},\Bar{b}}$.

The successful runs of the algorithm have an average convergence time of 2.5 hours.

\begin{figure}[tp]
    \centering
    \includegraphics[width=0.75\linewidth]{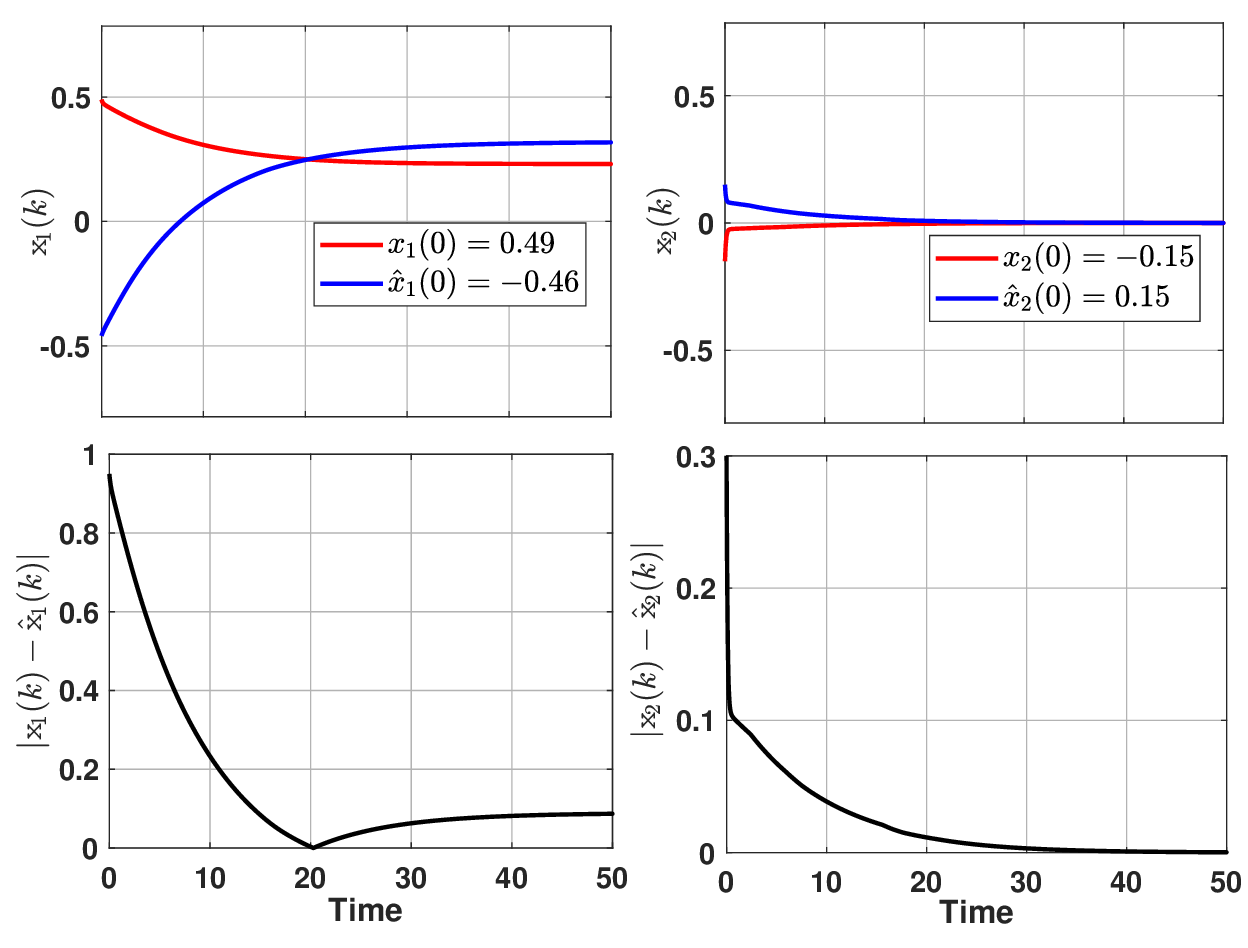}
    \caption{Top: Angular position (left) and velocity (right) of the manipulator, where the blue curve is influenced under input $\w(k) = 0.4 \in \W$, and the red curve is influenced under input $\w(k) = 0.3 \in \W$ for all $k \in \N_0$. Bottom: The difference in angular positions (left) and velocities (right) subjected to different initial conditions and input torques.}
    \label{fig:sim2}
\end{figure}

One can see from Figure \ref{fig:sim2} that under different input conditions, the trajectories corresponding to various states starting from different initial conditions maintain the same distance or converge to a particular trajectory after some time instances under the influence of the controller.

\subsection{Jet Engine Model}
We consider a nonlinear Moore-Grietzer Jet Engine Model in no-stall mode \cite{jet_engine}, whose discrete-time dynamics is governed by the following set of equations:
\begin{align}
    \mathsf{x}_1(k+1) &= \mathsf{x}_1(k) + \tau \big( - \mathsf{x}_2(k) - 1.5\mathsf{x}_1^2(k) - 0.5\mathsf{x}_1^3(k)\big),\notag \\
    \mathsf{x}_2(k+1) &= \mathsf{x}_2(k) + \tau (\mathsf{u}(k)),
\end{align}
where $\mathsf{x}_1 = \mu - 1,\mathsf{x}_2 = \zeta - \rho - 2$ with $\mu, \zeta, \rho$ denote the mass flow, the pressure rise and a constant, respectively. $\tau = 0.01$ is the sampling time. We consider the state space of the system to be $\X = [-0.25,0.25]\times[-0.25,0.25]$. Moreover, we consider the input set to be bounded within $\W = [-0.5,0.5]$. Also, we consider the model to be unknown. However, the Lipschitz constants $\mathcal{L}_x = 0.93, \mathcal{L}_u = 0.01$are known for the system. 

The goal is to synthesize a controller to enforce the system to be $\delta$-ISS. To do this, we first fix the training hyper-parameters as $\epsilon = 0.0157, \mathcal{L}_L = 1, \mathcal{L}_C = 10, k_1 = 0.00001, k_2 = 0.5, k_3 = 0.001, k_w = 0.01$. So, the Lipschitz constant $\mathcal{L}$ according to Theorem \ref{th:constr} is $3.351$. We fix the structure of $V_{\theta,b}$ as {$l_v = 1, h_v^1 = 40$} and $g_{\Bar{\theta},\Bar{b}}$ as $l_c = 1, h_c^1 = 15$. The training algorithm converges to obtain $\delta$-ISS-CLF $V_{\theta,b}$ along with $\eta = -0.0169$. Hence, $\eta+\mathcal{L}\epsilon = -0.0169 + 3.351\times0.005 = -0.000145$, therefore, using Theorem \ref{th:guarantee}, we can guarantee that the obtained $\delta$-ISS-CLF $V_{\theta,b}$ is valid and that the closed-loop system is assured to be incrementally input-to-state stable under the influence of the controller $g_{\Bar{\theta},\Bar{b}}$.

The successful runs of the algorithm have an average convergence time of 2.5 hours.

\begin{figure}[tp]
    \centering
    \includegraphics[width=0.75\linewidth]{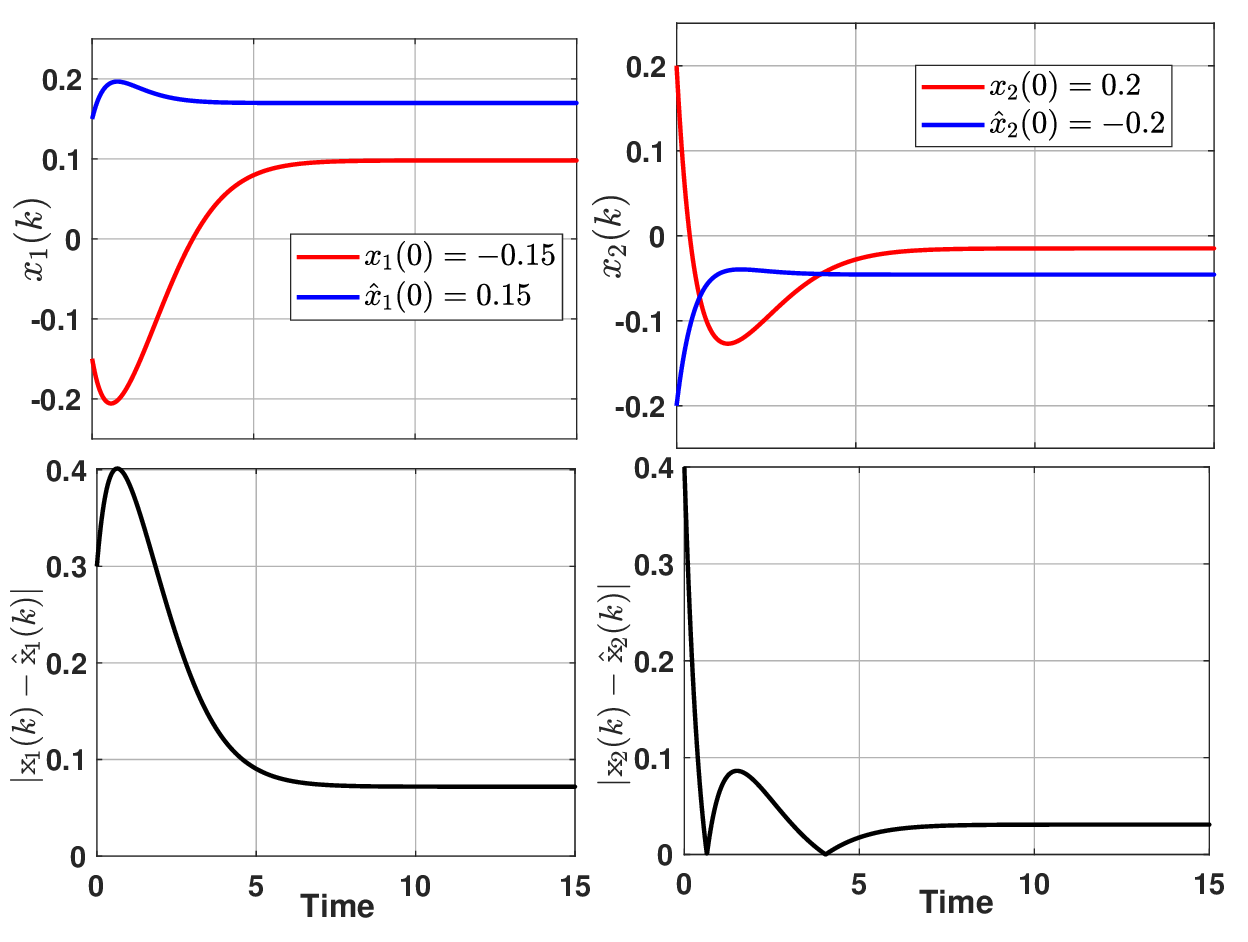}
    \caption{Top: Mass flow (left) and pressure rise (right) of the Jet-engine model, where the blue curve is influenced under input $\w(k) = -0.15 \in \W$, and the red curve is influenced under input $\w(k) = -0.3 \in \W$ for all $k \in \N_0$. Bottom: The difference in mass flow (left) and pressure rise (right) subjected to different initial conditions and input flows through the throttle.}
    \label{fig:sim3}
\end{figure}

As can be seen in Figure \ref{fig:sim3}, under different input conditions, the trajectories corresponding to various states starting from different initial conditions maintain the same distance after some time instances under the influence of the controller.

\begin{figure*}[t]
    \centering
    \begin{subfigure}{0.33\textwidth}
        \centering
        \includegraphics[width=0.95\textwidth]{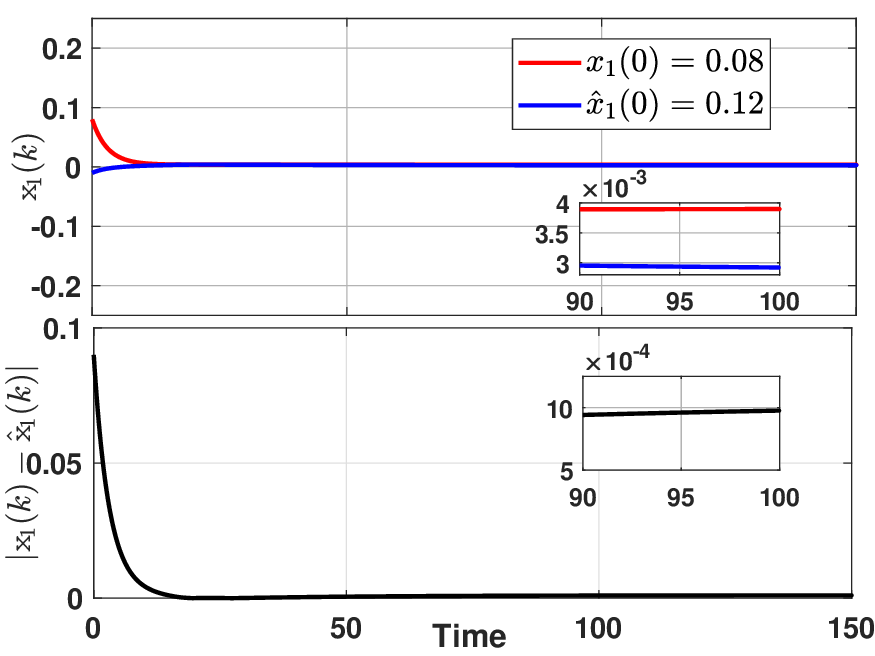}
        \caption{}
    \end{subfigure}
    \hfill
    \begin{subfigure}{0.33\textwidth}
        \centering
        \includegraphics[width=0.95\textwidth]{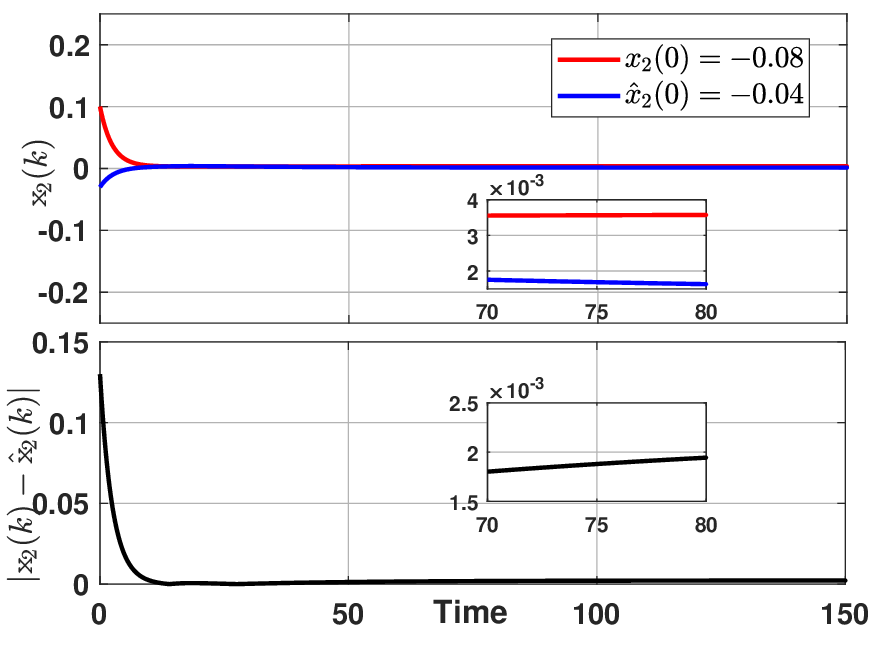}
        \caption{}
    \end{subfigure}
    \hfill
    \begin{subfigure}{0.33\textwidth}
        \centering
        \includegraphics[width=0.95\textwidth]{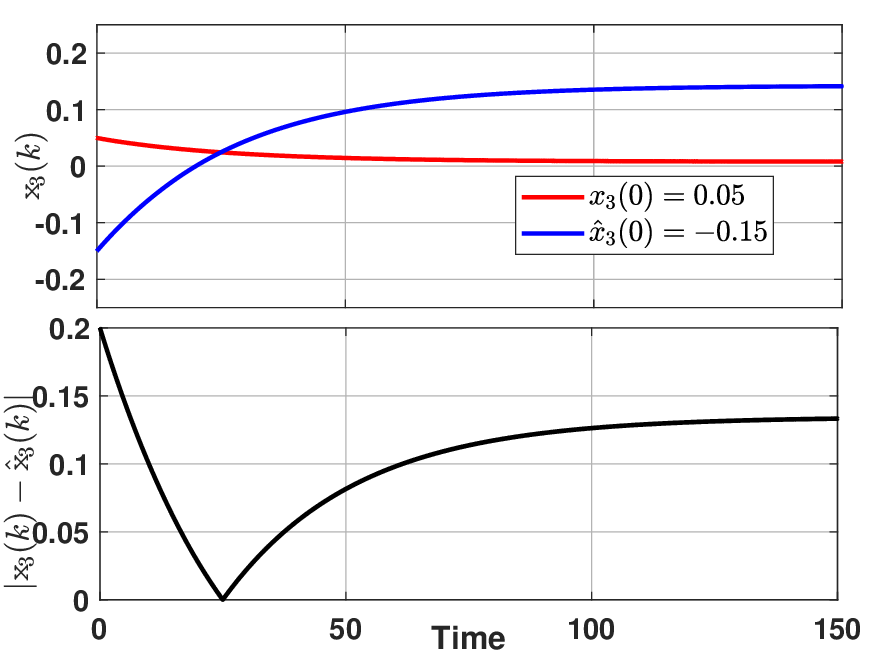}
        \caption{}
    \end{subfigure}
    \vspace{-0.5cm}
    \caption{Top: Trajectories ((a) $\omega_1$, (b) $\omega_2$, (c) $\omega_3$) starting from different initial conditions under different input signals (for all $k\in\N_0$, $\w(k)= -\sin(k)\in \W$ for blue curve and $\hat\w(k) = \cos^2(k/2) \in \W$ for red curve), Bottom: The difference between the trajectories corresponding to different states.}
    \label{fig:sim6}
\end{figure*}

\subsection{Rotating Spacecraft Model}
We consider another example of a rotating rigid spacecraft model \cite{khalil2002nonlinear}, whose discrete-time dynamics is governed by the following set of equations:
\begin{align}
    \mathsf{x}_1(k+1) &= \mathsf{x}_1(k) + \tau \Big( \frac{J_2 - J_3}{J_1} \mathsf{x}_2(k) \mathsf{x}_3(k) + \frac{1}{J_1} \mathsf{u}_1(k)\Big),\notag \\
    \mathsf{x}_2(k+1) &= \mathsf{x}_2(k) + \tau \Big(\frac{J_3 - J_1}{J_2} \mathsf{x}_1(k) \mathsf{x}_3(k) + \frac{1}{J_2} \mathsf{u}_2(k)\Big),\notag \\
     \mathsf{x}_3(k+1) &= \mathsf{x}_3(k) + \tau \Big(\frac{J_1 - J_2}{J_3} \mathsf{x}_1(k) \mathsf{x}_2(k) + \frac{1}{J_3} \mathsf{u}_3(k)\Big),
\end{align}
where $\mathsf{x} = [\mathsf{x}_1, \mathsf{x}_2, \mathsf{x}_3]^\top$ denotes angular velocities $\omega_1, \omega_2, \omega_3$ along the principal axes respectively,  $\mathsf{u} = [\mathsf{u}_1, \mathsf{u}_2, \mathsf{u}_3]^\top$ represents the torque input, and with $J_1 = 200, J_2 = 200, J_3 = 100$ denote the principal moments of inertia. $\tau = 0.01$ is the sampling time. We consider the state space of the system to be $\X = [-0.25,0.25]\times[-0.25,0.25]\times [-0.25,0.25]$. Moreover, we consider the external input set to be bounded within $\W = [-1,1]$. Note that the internal input is of 3 dimensions while the external input is of a single dimension, \textit{i.e.,} $\U \in \R^3$, but $\W \in \R$. Also, we consider the model to be unknown. { However, the Lipschitz constants $\mathcal{L}_x = 1, \mathcal{L}_u = 0.01$ are known for the system}.

We fix the training hyper-parameters as $\epsilon = 0.0125, \mathcal{L}_L = 1, \mathcal{L}_C = 40, {k = [0.00001, 0.5, 0.001, 0.01], \Gamma = [2,2,2,2]}$. So, the Lipschitz constant according to Theorem \ref{th:constr} is $3.651$. We fix the structure of $V_{\theta,b}$ as {$l_v = 1, h_v^1 = 60$} and $g_{\Bar{\theta},\Bar{b}}$ as $l_c = 1, h_c^1 = 40$. {The activation function for both $\delta$-ISS-CLF and the controller is considered to be `ReLU'.} The training algorithm converges to obtain $\delta$-ISS-CLF $V_{\theta,b}$ along with $\eta = -0.0460$. Hence, $\eta+\mathcal{L}\epsilon = -0.0460 + 3.651\times0.0125 = -0.00036$, therefore, using Theorem \ref{th:guarantee}, we can guarantee that the obtained $\delta$-ISS-CLF $V_{\theta,b}$ is valid and that the closed-loop system is assured to be incrementally input-to-state stable under the influence of the controller $g_{\Bar{\theta},\Bar{b}}$.

The successful runs of the algorithm {including the dataset generation time} have an average convergence time of 5.5 hours. {Note that the computational complexity grows rapidly with respect to system dimension as proposed in \cite{esfahani2014performance}}.

One can see from Figure \ref{fig:sim6}(a), \ref{fig:sim6}(b), and \ref{fig:sim6}(c), under the different input conditions, the trajectories corresponding to various states starting from different initial conditions maintain the same distance after some time instances under the influence of the controller.

\subsection{Discussion}\label{subsec:Discuss}
The controller synthesis technique proposed in this work for achieving incremental input-to-state stability in closed-loop systems offers several advantages over existing methods. As the current technique relies on synthesizing the controller using data of the system, hence designing the controller does not require the system to be either fully actuated or affine structure as done in previous conventional techniques \cite{zamani2013backstepping, jagtap2017backstepping}. This neural network based controller synthesis is able to make underactuated systems incrementally stable without the knowledge of dynamics. For instance, our neural network-based controller successfully ensures incremental stability in the one-link manipulator and jet-engine model without requiring detailed dynamic equations \cite{basu2025controller}. The approach is also reliable for the case of nonaffine systems as well, as seen in the scalar example.

{Also, from the final case study, although the system has three independent control inputs, the neural controller can generate them using the system's state and a single external input $w \in \W \subset \R$, so the closed-loop system becomes $\delta$-ISS with respect to a scalar exogenous input, which significantly simplifies control implementation.} Now, as seen from \cite{zamani2017towards}, the computation complexity for the abstraction of an incrementally stable system lies in the dimension of the input to the system. As we have made the system incrementally stable with one input only, the computational complexity of the abstraction will reduce drastically once the controller is trained. This is helpful in finding the scalable symbolic abstraction of the system and, thus, will enable the efficient design of another formally verified controller to perform other specifications for the closed-loop system. 
%%%%%%%%%%%%%%%%%%%%%%%%%%%%%%%%%%%%%%%%%%%%%%%%%%%%%%%%%%%%%%%%%%%%%%%%%%%%%%%%
%%--------------------------------NEW SECTION---------------------------------%%
%%%%%%%%%%%%%%%%%%%%%%%%%%%%%%%%%%%%%%%%%%%%%%%%%%%%%%%%%%%%%%%%%%%%%%%%%%%%%%%%
\section{Conclusion and Future Work}\label{sec:Conclusion}
In this paper, we presented a new training approach for unknown discrete-time systems that simultaneously synthesizes a controller and a verifiably correct incremental input-to-state stable control Lyapunov function, both parameterized as neural networks. The notion of incremental input-to-state state control Lyapunov function was introduced for the very first time. We highlighted the crucial challenges that are faced while solving an ROP to ensure the existence of $\delta$-ISS-CLF and thereby mapped the problem into SOP. Consequently, we presented a validity condition, by virtue of which solving SOP ensures satisfying the corresponding ROP, necessitating the existence of an appropriate $\delta$-ISS-CLF. We illustrated various case studies in support of our proposed claims. A promising future direction is to extend the framework to construct $\delta$-ISS-CLF for continuous-time systems. Moreover, we plan to extend these results to guarantee incremental ISS for different systems in stochastic settings {where stochasticity might occur due to measurement noise} and large-scale interconnected systems.

% \section*{References}
\bibliographystyle{ieeetr} % We choose the "plain" reference style
\bibliography{sources} % Entries are in the refs.bib file

\begin{thebibliography}{10}

\bibitem{analog}
B.~N. Bond, Z.~Mahmood, Y.~Li, R.~Sredojevic, A.~Megretski, V.~Stojanovi, Y.~Avniel, and L.~Daniel, ``Compact modeling of nonlinear analog circuits using system identification via semidefinite programming and incremental stability certification,'' {\em IEEE Transactions on Computer-Aided Design of Integrated Circuits and Systems}, vol.~29, no.~8, pp.~1149--1162, 2010.

\bibitem{cyclic_feedback}
A.~Hamadeh, G.-B. Stan, R.~Sepulchre, and J.~Goncalves, ``Global state synchronization in networks of cyclic feedback systems,'' {\em IEEE Transactions on Automatic Control}, vol.~57, no.~2, pp.~478--483, 2012.

\bibitem{symbolic1}
G.~Pola, A.~Girard, and P.~Tabuada, ``Approximately bisimilar symbolic models for nonlinear control systems,'' {\em Automatica}, vol.~44, no.~10, pp.~2508--2516, 2008.

\bibitem{zamani2017towards}
M.~Zamani, I.~Tkachev, and A.~Abate, ``Towards scalable synthesis of stochastic control systems,'' {\em Discrete Event Dynamic Systems}, vol.~27, pp.~341--369, 2017.

\bibitem{jagtap2020symbolic}
P.~Jagtap and M.~Zamani, ``Symbolic models for retarded jump--diffusion systems,'' {\em Automatica}, vol.~111, p.~108666, 2020.

\bibitem{jagtap2017quest}
P.~Jagtap and M.~Zamani, ``{QUEST}: A tool for state-space quantization-free synthesis of symbolic controllers,'' in {\em International conference on quantitative evaluation of systems}, pp.~309--313, Springer, 2017.

\bibitem{inter_osci}
G.-B. Stan and R.~Sepulchre, ``Analysis of interconnected oscillators by dissipativity theory,'' {\em IEEE Transactions on Automatic Control}, vol.~52, no.~2, pp.~256--270, 2007.

\bibitem{synchComplex}
G.~Russo and M.~di~Bernardo, ``Contraction theory and master stability function: Linking two approaches to study synchronization of complex networks,'' {\em IEEE Transactions on Circuits and Systems II: Express Briefs}, vol.~56, no.~2, pp.~177--181, 2009.

\bibitem{Contraction}
W.~Lohmiller and J.~E. Slotine, ``On contraction analysis for non-linear systems,'' {\em Automatica}, vol.~34, no.~6, pp.~683--696, 1998.

\bibitem{Conv_dyn}
A.~Pavlov, N.~Van De~Wouw, and H.~Nijmeijer, {\em Uniform output regulation of nonlinear systems: a convergent dynamics approach}, vol.~205.
\newblock Springer, 2006.

\bibitem{angeli2002lyapunov}
D.~Angeli, ``A {L}yapunov approach to incremental stability properties,'' {\em IEEE Transactions on Automatic Control}, vol.~47, no.~3, pp.~410--421, 2002.

\bibitem{DT-ISS_prop}
D.~N. Tran, B.~S. R{\"u}ffer, and C.~M. Kellett, ``Incremental stability properties for discrete-time systems,'' in {\em 2016 IEEE 55th Conference on Decision and Control (CDC)}, pp.~477--482, IEEE, 2016.

\bibitem{zamani_inc}
M.~Zamani and R.~Majumdar, ``A {L}yapunov approach in incremental stability,'' in {\em 2011 50th IEEE conference on decision and control and European control conference}, pp.~302--307, IEEE, 2011.

\bibitem{biemond2018incremental}
J.~B. Biemond, R.~Postoyan, W.~M.~H. Heemels, and N.~Van De~Wouw, ``Incremental stability of hybrid dynamical systems,'' {\em IEEE transactions on automatic control}, vol.~63, no.~12, pp.~4094--4109, 2018.

\bibitem{chaillet2013razumikhin}
A.~Chaillet, A.~Y. Pogromsky, and B.~S. R{\"u}ffer, ``A {R}azumikhin approach for the incremental stability of delayed nonlinear systems,'' in {\em 52nd IEEE Conference on Decision and Control}, pp.~1596--1601, IEEE, 2013.

\bibitem{dey2023incremental}
B.~S. Dey, I.~N. Kar, and P.~Jagtap, ``On incremental stability of interconnected switched systems,'' {\em arXiv preprint arXiv:2308.12746}, 2023.

\bibitem{zamani2011backstepping}
M.~Zamani and P.~Tabuada, ``Backstepping design for incremental stability,'' {\em IEEE Transactions on Automatic Control}, vol.~56, no.~9, pp.~2184--2189, 2011.

\bibitem{zamani2013backstepping}
M.~Zamani, N.~van~de Wouw, and R.~Majumdar, ``Backstepping controller synthesis and characterizations of incremental stability,'' {\em Systems \& Control Letters}, vol.~62, no.~10, pp.~949--962, 2013.

\bibitem{jagtap2017backstepping}
P.~Jagtap and M.~Zamani, ``Backstepping design for incremental stability of stochastic {H}amiltonian systems with jumps,'' {\em IEEE Transactions on Automatic Control}, vol.~63, no.~1, pp.~255--261, 2017.

\bibitem{sangeerth2025controller}
P.~Sangeerth, D.~S. Sundarsingh, B.~S. Dey, and P.~Jagtap, ``Controller for incremental input-to-state practical stabilization of partially unknown systems with invariance guarantees,'' {\em arXiv preprint arXiv:2510.10450}, 2025.

\bibitem{zaker2024controller}
M.~Zaker, D.~Angeli, and A.~Lavaei, ``Certified learning of incremental {ISS} controllers for unknown nonlinear polynomial dynamics,'' {\em arXiv preprint arXiv:2412.03901}, 2024.

\bibitem{formal_nn_Lyapunov}
A.~Abate, D.~Ahmed, M.~Giacobbe, and A.~Peruffo, ``Formal synthesis of {L}yapunov neural networks,'' {\em IEEE Control Systems Letters}, vol.~5, no.~3, pp.~773--778, 2020.

\bibitem{Neural_CBF}
M.~Anand and M.~Zamani, ``Formally verified neural network control barrier certificates for unknown systems,'' {\em IFAC-PapersOnLine}, vol.~56, no.~2, pp.~2431--2436, 2023.
\newblock 22nd IFAC World Congress.

\bibitem{tayal2024learning}
M.~Tayal, H.~Zhang, P.~Jagtap, A.~Clark, and S.~Kolathaya, ``Learning a formally verified control barrier function in stochastic environment,'' {\em arXiv preprint arXiv:2403.19332}, 2024.

\bibitem{FV_DD}
A.~Nejati, A.~Lavaei, P.~Jagtap, S.~Soudjani, and M.~Zamani, ``Formal verification of unknown discrete- and continuous-time systems: {A} data-driven approach,'' {\em IEEE Transactions on Automatic Control}, vol.~68, no.~5, pp.~3011--3024, 2023.

\bibitem{chang2019neural}
Y.-C. Chang, N.~Roohi, and S.~Gao, ``Neural lyapunov control,'' {\em Advances in neural information processing systems}, vol.~32, 2019.

\bibitem{yang2024lyapunov}
L.~Yang, H.~Dai, Z.~Shi, C.-J. Hsieh, R.~Tedrake, and H.~Zhang, ``Lyapunov-stable neural control for state and output feedback: A novel formulation,'' {\em arXiv preprint arXiv:2404.07956}, 2024.

\bibitem{abate2020formal}
A.~Abate, D.~Ahmed, M.~Giacobbe, and A.~Peruffo, ``Formal synthesis of lyapunov neural networks,'' {\em IEEE Control Systems Letters}, vol.~5, no.~3, pp.~773--778, 2020.

\bibitem{basu2025formally}
A.~Basu, B.~S. Dey, and P.~Jagtap, ``Formally verified neural {L}yapunov function for incremental input-to-state stability of unknown systems,'' {\em arXiv preprint arXiv:2501.05778}, 2025.

\bibitem{DT-ISS}
F.~Bayer, M.~Bürger, and F.~Allgöwer, ``{Discrete-time incremental {ISS}: A framework for Robust NMPC},'' in {\em 2013 European Control Conference (ECC)}, pp.~2068--2073, 2013.

\bibitem{liu2019compositional}
S.~Liu and M.~Zamani, ``Compositional synthesis of almost maximally permissible safety controllers,'' in {\em 2019 American Control Conference (ACC)}, pp.~1678--1683, IEEE, 2019.

\bibitem{esfahani2014performance}
P.~M. Esfahani, T.~Sutter, and J.~Lygeros, ``Performance bounds for the scenario approach and an extension to a class of non-convex programs,'' {\em IEEE Transactions on Automatic Control}, vol.~60, no.~1, pp.~46--58, 2014.

\bibitem{pauli2022b}
P.~Pauli, A.~Koch, J.~Berberich, P.~Kohler, and F.~Allgöwer, ``Training robust neural networks using {L}ipschitz bounds,'' {\em IEEE Control Systems Letters}, vol.~6, pp.~121--126, 2022.

\bibitem{Lipschitz}
G.~R. Wood and B.~P. Zhang, ``Estimation of the {L}ipschitz constant of a function,'' {\em Journal of Global Optimization}, vol.~8, pp.~91--103, 1996.

\bibitem{ruder2016}
S.~Ruder, ``An overview of gradient descent optimization algorithms,'' {\em arXiv preprint arXiv:1609.04747}, 2016.

\bibitem{zhao2021learning}
H.~Zhao, X.~Zeng, T.~Chen, Z.~Liu, and J.~Woodcock, ``Learning safe neural network controllers with barrier certificates,'' {\em Formal Aspects of Computing}, vol.~33, no.~3, pp.~437--455, 2021.

\bibitem{nn_lr}
Y.~Li, C.~Wei, and T.~Ma, ``Towards explaining the regularization effect of initial large learning rate in training neural networks,'' {\em Advances in neural information processing systems}, vol.~32, 2019.

\bibitem{lewis2020neural}
F.~Lewis, S.~Jagannathan, and A.~Yesildirak, {\em Neural network control of robot manipulators and non-linear systems}.
\newblock CRC press, 2020.

\bibitem{murray2017mathematical}
R.~M. Murray, Z.~Li, and S.~S. Sastry, {\em A mathematical introduction to robotic manipulation}.
\newblock CRC press, 2017.

\bibitem{jet_engine}
M.~Krstic and P.~V. Kokotovic, ``Lean backstepping design for a jet engine compressor model,'' in {\em Proceedings of International Conference on Control Applications}, pp.~1047--1052, IEEE, 1995.

\bibitem{khalil2002nonlinear}
H.~Khalil, {\em Nonlinear Systems}.
\newblock Pearson Education, Prentice Hall, 2002.

\bibitem{basu2025controller}
A.~Basu, B.~S. Dey, and P.~Jagtap, ``Formally verified neural network controllers for incremental input-to-state stability of unknown discrete-time systems,'' {\em arXiv preprint arXiv:2503.04129}, 2025.

\bibitem{jiang2001input}
Z.-P. Jiang and Y.~Wang, ``Input-to-state stability for discrete-time nonlinear systems,'' {\em Automatica}, vol.~37, no.~6, pp.~857--869, 2001.

\bibitem{jiang2002converse}
Z.-P. Jiang and Y.~Wang, ``A converse {L}yapunov theorem for discrete-time systems with disturbances,'' {\em Systems \& control letters}, vol.~45, no.~1, pp.~49--58, 2002.

\end{thebibliography}

\subsection*{Appendix A: Proof of Theorem \ref{th:admit}} \label{appendix:admit}
The proof follows a similar approach of \cite[Theorem 8]{DT-ISS_prop}. Let the closed-loop system admit a $\delta$-ISS-CLF satisfying the conditions of Definition \ref{def:ISS-Lf}. The lower bound (i) and inequality (ii) imply: $\forall x, \hat{x} \in \X, w, \hat{w} \in \W$:
\begin{align}\label{eq:i+ii}
    & V(f(x, g(x,w)),f(\hat{x},g(\hat{x},\hat{w}))) - V(x,\hat{x}) \leq -\alpha(V(x,\hat{x})) + \sigma(|w-\hat{w}|),
\end{align}
where $\alpha \in \mathcal{K}_\infty$ is defined as $\alpha(s) := \alpha_3 \circ \alpha_1^{-1}(s)$ for all $s \in \R_0^+$.

We denote $id$ as an identity function, i.e. $id(s) := s$ for all $ s \in \R_0^+$. Assume $\chi$ be any class $\mathcal{K}_{\infty}$ such that $id - \chi \in \mathcal{K}_{\infty}$. Without loss of generality, we assume $id - \alpha \in \mathcal{K}$ \cite[Lemma B.1]{jiang2001input}.

Let $\w, \hat{\w}$ denote two input sequences such that $\w, \hat{\w}: \N_0 \rightarrow \W$. Now, define the set :
\begin{align}\label{set:S}
    \mathbb{S}: &= \{(\epsilon_1, \epsilon_2)| \epsilon_1, \epsilon_2 \in \X, V(\epsilon_1, \epsilon_2) \leq \alpha^{-1} \circ \chi^{-1} \circ \sigma(\lVert \w - \hat{\w} \rVert) \}.
\end{align}

\begin{lemma}
    The set $\mathbb{S}$ is forward invariant.
\end{lemma}

\begin{proof}
    Let $(x, \hat{x}) \in \mathbb{S}$. Also, $\w(0) = w, \hat{\w}(0) = \hat{w} \in \W$ be the first elements of $\w, \hat{\w}$ respectively. Note that the set $\X$ is compact and hence under the controller $g$, it is control forward invariant. Therefore, the terms $f(x, g(x,w))$ and $f(\hat{x}, g(\hat{x}, \hat{w})) \in \X$.
    
    Then, $V(x, \hat{x}) \leq \alpha^{-1} \circ \chi^{-1} \circ \sigma(\lVert \w - \hat{\w} \rVert)$ and combining with \eqref{eq:i+ii}, we have:
    \begin{align*}
        &V(f(x, g(x,w)), f(\hat{x}, g(\hat{x}, \hat{w})))  \leq V(x, \hat{x}) - \alpha(V(x, \hat{x})) + \sigma(\lVert \w - \hat{\w} \rVert) \notag \\
        & \leq (id - \alpha)V(x, \hat{x}) + \sigma(\lVert \w - \hat{\w} \rVert)\leq (id - \alpha) \circ \alpha^{-1} \circ \chi^{-1} \circ \sigma(\lVert \w - \hat{\w} \rVert) + \sigma(\lVert \w - \hat{\w} \rVert) \notag \\
        & \leq \alpha^{-1} \circ \chi^{-1} \circ \sigma(\lVert \w - \hat{\w} \rVert) - (id - \chi) \circ \chi^{-1} \circ \sigma(\lVert \w - \hat{\w} \rVert).
    \end{align*}
    Since $(id - \chi) \in \mathcal{K}_{\infty}$, hence $- (id - \chi) \circ \chi^{-1} \circ \sigma(\lVert \w - \hat{\w} \rVert) \leq 0$. So, essentially, 
    \begin{align*}
        V(f(x, g(x,w)), f(\hat{x}, g(\hat{x}, \hat{w}))) \leq \alpha^{-1} \circ \chi^{-1} \circ \sigma(\lVert \w - \hat{\w} \rVert)
    \end{align*}
    which implies that $(f(x, g(x,w)), f(\hat{x}, g(\hat{x}, \hat{w}))) \in \mathbb{S}$. Hence, we can say that the set $\mathbb{S}$ is forward invariant.
\end{proof}

Let us denote the state of the closed-loop system at time instance $k$ under the action of input sequence $\mathsf{w}$ as $\mathsf{x}(k,x,g(x,\mathsf{w}(k)))$ starting from the initial condition $x$. 

\textit{Part-I}: For $(x, \hat{x}) \in \mathbb{S}$, applying condition (i) of Definition \ref{def:ISS-Lf}, we obtain $
    \alpha_1(|\mathsf{x}(k,x,g(x,\mathsf{w}(k)) - \mathsf{x}(k,\hat{x},g(\hat{x},\mathsf{\hat{w}}(k)))|)
    \leq V(\mathsf{x}(k,x,g(x,\mathsf{w}(k))), \mathsf{x}(k,\hat{x},g(\hat{x},\mathsf{\hat{w}}(k))))
    \leq \alpha^{-1} \circ \chi^{-1} \circ \sigma(\lVert \w - \hat{\w} \rVert).$
Therefore, 
\begin{align}\label{eq:cl_gamma}
    & |\mathsf{x}(k,x,g(x,\mathsf{w}(k))) - \mathsf{x}(k,\hat{x},g(\hat{x},\mathsf{\hat{w}}(k)))| \leq \alpha_1^{-1} \circ \alpha^{-1} \circ \chi^{-1} \circ \sigma(\lVert \w - \hat{\w} \rVert) =: \bar{\gamma}(\lVert \w - \hat{\w} \rVert),
\end{align}
where $\bar{\gamma}(s) := \alpha_1^{-1} \circ \alpha^{-1} \circ \chi^{-1} \circ \sigma(s)$ for all $s \in \R_0^+$ and $\bar{\gamma}\in \mathcal{K}_\infty$.

\textit{Part-II}: For $(x, \hat{x}) \notin \mathbb{S}$, \eqref{set:S} shows that $\chi \circ \alpha (V(x, \hat{x})) > \sigma(\lVert \w - \hat{\w} \rVert)$. Hence, \eqref{eq:i+ii} gives
\begin{align*}
    & V(f(x, g(x,w)), f(\hat{x}, g(\hat{x}, \hat{w}))) - V(x, \hat{x}) \leq -\alpha(V(x, \hat{x})) + \chi \circ \alpha(V(x,\hat{x})) \leq(id - \chi)\circ\alpha(V(x,\hat{x})).
\end{align*}
Since, $id - \chi \in \mathcal{K}_{\infty}$, using the comparison lemma \cite[Lemma 4.3]{jiang2002converse}, we infer there exists some $\Tilde{\beta} \in \mathcal{KL}$ so that, 
\begin{align*}
    \alpha_1(|\mathsf{x}(k,x,g(x,\mathsf{w}(k))) - \mathsf{x}(k,\hat{x},g(\hat{x},\mathsf{\hat{w}}(k)))|) &\leq V(\mathsf{x}(k,x,g(x,\mathsf{w}(k))), \mathsf{x}(k,\hat{x},g(\hat{x},\mathsf{\hat{w}}(k)))) \notag \\
    & \leq \Tilde{\beta}(V(x, \hat{x}),k) \leq \Tilde{\beta}(\alpha_2(|x - \hat{x}|),k),
\end{align*}
which essentially implies
\begin{align}\label{eq:cl_beta}
    &|\mathsf{x}(k,x,g(x,\mathsf{w}(k))) - \mathsf{x}(k,\hat{x},g(\hat{x},\mathsf{\hat{w}}(k)))| \leq \alpha_1^{-1} \circ \tilde{\beta}(\alpha_2(|x - \hat{x}|),k) =: \bar{\beta}((|x - \hat{x}|),k),
\end{align}
where $\bar{\beta}(r,s) := \alpha_1^{-1} \circ \Tilde{\beta}(\alpha_2(r),s)$ for all $r,s \in \R_0^+$ and $\bar{\beta} \in \mathcal{KL}$. 

Now, combining \eqref{eq:cl_gamma} and \eqref{eq:cl_beta}, we obtain
\begin{align*}
    |\mathsf{x}(k,x,g(x,\mathsf{w}(k))) - \mathsf{x}(k,\hat{x},g(\hat{x},\mathsf{\hat{w}}(k)))| &\leq  \max\{\bar{\beta}((|x - \hat{x}|),k), \bar{\gamma}(\lVert \w - \hat{\w} \rVert)\} \notag \\
    & \leq \bar{\beta}((|x - \hat{x}|),k) + \bar{\gamma}(\lVert \w - \hat{\w} \rVert).
\end{align*}
So, the closed-loop system that satisfies the $\delta$-ISS-CLF condition under the action of the controller is incrementally ISS.

\subsection*{Appendix B: Proof of Theorem \ref{th:constr}} \label{appendix:constr}
Here, we demonstrate under condition \eqref{eq:cond}, the obtained $\delta$-ISS-CLF and the controller from SCP satisfies the conditions of Definition \ref{def:ISS-Lf}.  The optimal $\eta_S^*$, obtained through solving the \eqref{eq:SCP}, guarantees for any $x_q,x_r \in \mathcal{X}, w_q, w_r \in \Wbo$ we have: 
\begin{align*}
& -V_{\theta,b}(x_q,x_r) + k_1|x_q-x_r|^{\gamma_1} \leq \eta_S^*,\\
& V_{\theta,b}(x_q,x_r) - k_2|x_q-x_r|^{\gamma_2} \leq \eta_S^*, \\ 
& V_{\theta,b}(f(x_q, g_{\bar{\theta}, \bar{b}}(x_q,w_q)),f(x_r, g_{\bar{\theta}, \bar{b}}(x_r,w_r))) - V_{\theta,b}(x_q,x_r) + k_3 |x_q - x_r|^{\gamma_3}  - k_w|w_q - w_r|^{\gamma_w}\leq \eta_S^*, \\
& h(f(x_q, g(x_q,w_q))) - h(x_q) \leq \eta_S^*.
\end{align*}
Now from \eqref{set:SCP}, we infer that $\forall x \in \X, \text{there exists}$ $ x_r $ s.t. $|x-x_r| \leq \varepsilon$ as well as $\forall w \in \W, \text{there exists}$ $ w_r $ s.t. $|w-w_r| \leq \varepsilon$. Hence, $\forall x,\hat{x} \in \X, \forall w, \hat{w} \in \W$:
\begin{align*}
    &\text{(a)} -V_{\theta,b}(x,\hat{x}) + k_1|x-\hat{x}|^{\gamma_1} \\
    & = \big(-V_{\theta,b}(x,\hat{x}) + V_{\theta,b}(x_q,x_r)\big) + \big( - V_{\theta,b}(x_q,x_r)
    + k_1|x_q-x_r|^{\gamma_1}\big) + \big(- k_1|x_q-x_r|^{\gamma_1} + k_1|x-\hat{x}|^{\gamma_1}\big) \\
    & \leq \mathcal{L}_L |(x,\hat{x}) - (x_q,x_r)| + \eta_S^* + 2\mathcal{L}_1\varepsilon \leq \sqrt{2}\mathcal{L}_L\varepsilon + \eta_S^* + 2\mathcal{L}_1\varepsilon \leq \mathcal{L}\varepsilon + \eta_S^* \leq 0. \\
    &\text{(b)} V_{\theta,b}(x,\hat{x}) - k_2|x-\hat{x}|^{\gamma_2} \\
    & = \big(V_{\theta,b}(x,\hat{x}) - V_{\theta,b}(x_q,x_r)\big) + \big(  V_{\theta,b}(x_q,x_r) - k_2|x_q-x_r|^{\gamma_2}\big) + \big(k_2|x_q-x_r|^{\gamma_2} - k_2|x-\hat{x}|^{\gamma_2}\big) \\
    & \leq \mathcal{L}_L |(x,\hat{x}) - (x_q,x_r)| + \eta_S^* + 2\mathcal{L}_2\varepsilon \leq \sqrt{2}\mathcal{L}_L\varepsilon + \eta_S^* + 2\mathcal{L}_2\varepsilon \leq \mathcal{L}\varepsilon + \eta_S^* \leq 0. \\
    &\text{(c)} V_{\theta,b}(f(x, g_{\bar{\theta}, \bar{b}}(x,w)),f(\hat{x},g_{\bar{\theta}, \bar{b}}(\hat{x}, \hat{w}))) - V_{\theta,b}(x,\hat{x}) + k_3|x - \hat{x}|^{\gamma_3} - k_w|w - \hat{w}|^{\gamma_w}\\
     & = V_{\theta,b}(f(x, g_{\bar{\theta}, \bar{b}}(x,w)),f(\hat{x},g_{\bar{\theta}, \bar{b}}(\hat{x}, \hat{w})) - V_{\theta,b}(x,\hat{x}) - V_{\theta,b}(f(x_q, g_{\bar{\theta}, \bar{b}}(x_q,w_q)),f(x_r,g_{\bar{\theta}, \bar{b}}(x_r, w_r))) \\
     & \quad {+} V_{\theta,b}(f(x_q, g_{\bar{\theta}, \bar{b}}(x_q,w_q)),f(x_r,g_{\bar{\theta}, \bar{b}}(x_r, w_r))) + V_{\theta,b}(x_q,x_r) - V_{\theta,b}(x_q,x_r) + k_3|x - \hat{x}|^{\gamma_3} \\
     & \quad + k_3|x_q - x_r|^{\gamma_3} - k_3|x_q -x_r|^{\gamma_3} - k_w|w - \hat{w}|^{\gamma_w} - k_w|w_q - w_r|^{\gamma_w} + k_w|w_q -w_r|^{\gamma_w} \\
     & \leq \eta_S^* + \mathcal{L}_L|\big(f(x, g_{\bar{\theta}, \bar{b}}(x,w)),f(\hat{x},g_{\bar{\theta}, \bar{b}}(\hat{x}, \hat{w}))\big)- \big(f(x_q, g_{\bar{\theta}, \bar{b}}(x_q,w_q)),f(x_r,g_{\bar{\theta}, \bar{b}}(x_r, w_r))\big)| + \sqrt{2}\mathcal{L}_L\varepsilon \\
     & \quad + 2(\mathcal{L}_3+\mathcal{L}_w)\varepsilon \\
     & \leq \eta_S^* + \sqrt{2}\mathcal{L}_L(\mathcal{L}_x + \sqrt{2}\mathcal{L}_u\mathcal{L}_C) \varepsilon + \sqrt{2}\mathcal{L}_L\varepsilon + 2(\mathcal{L}_3+\mathcal{L}_w)\varepsilon \\
     & \leq \eta_S^* + \big(\sqrt{2}\mathcal{L}_L (\mathcal{L}_x + \sqrt{2}\mathcal{L}_u\mathcal{L}_C + 1)+ 2(\mathcal{L}_3+\mathcal{L}_w )\big)\varepsilon \leq \eta_S^* + \mathcal{L}\varepsilon \leq 0. \\
     &\text{(d)} h(f(x,g_{\bar{\theta}, \bar{b}}(x,w))) - h(x) \\
     & = h(f(x,g_{\bar{\theta}, \bar{b}}(x,w))) - h(f(x_q,g_{\bar{\theta}, \bar{b}}(x_q,w_q))) + h(f(x_q,g_{\bar{\theta}, \bar{b}}(x_q,w_q))) - h(x) + h(x_q) - h(x_q) \\
     & \leq \mathcal{L}_h(f(x,g_{\bar{\theta}, \bar{b}}(x,w)) - f(x_q,g_{\bar{\theta}, \bar{b}}(x_q,w_q))) + \eta_S^* + \mathcal{L}_h \varepsilon \\
     & \leq \mathcal{L}_h(\mathcal{L}_x + \sqrt{2}\mathcal{L}_u \mathcal{L}_C)\varepsilon + \eta_S^* + \mathcal{L}_h \varepsilon \leq \eta_S^* + \mathcal{L}_h(\mathcal{L}_x + \sqrt{2}\mathcal{L}_u \mathcal{L}_C + 1)\varepsilon \leq \eta_S^* + \mathcal{L}\varepsilon \leq 0.
 \end{align*}
 
 Therefore, if the condition \eqref{eq:cond} is satisfied, the neural $\delta$-ISS-CLF will ensure the system to be $\delta$-ISS under the controller. 

\end{document}